\documentclass[aps,showpacs,preprintnumbers,twocolumn,superscriptaddress,floatfix,nofootinbib]{revtex4-2}

\usepackage[linesnumbered,lined,boxed,commentsnumbered,ruled,vlined]{algorithm2e}
\usepackage{algpseudocode}
\usepackage{amsfonts}
\usepackage{amsmath}
\usepackage{amssymb}
\usepackage{graphicx}
\usepackage{dcolumn}
\usepackage{bm,bbm}
\usepackage{xkcdcolors}
\usepackage{tabularx}
\usepackage{epstopdf}
\usepackage{xcolor}
\usepackage{mathrsfs}
\usepackage{subfigure}
\usepackage{caption}
\usepackage{mathtools}
\usepackage{float}
\usepackage{verbatim}
\usepackage{comment}
\usepackage{ragged2e}
\usepackage{physics}
\usepackage{hyperref}
\usepackage{url}
\usepackage{cleveref}
\usepackage{tikz}
\usepackage{utfsym}
\usepackage{multirow}
\usepackage{scalerel}
\usepackage{amsthm}
\usepackage{booktabs}
\usepackage[normalem]{ulem}

\DeclareCaptionJustification{justified}{\justifying}
\hypersetup{colorlinks=true, citecolor=orange, urlcolor=blue, linkcolor=magenta}

\definecolor{cel}{rgb}{0.0,0.53,0.74}
\definecolor{green}{rgb}{0.0,0.5,0.0}

\definecolor{colSDRG}{RGB}{51,117,56}
\definecolor{colSDCM}{RGB}{194,106,119}
\definecolor{colRec}{RGB}{230,158,0}
\definecolor{colpp}{RGB}{87,181,232}
\definecolor{colnn}{RGB}{0,158,115}
\definecolor{colpn}{RGB}{204,120,166}

\begin{document}

\title{Reproducing the first and second moments of empirical degree distributions}

\author{Mattia Marzi}
\email{mattia.marzi@imtlucca.it}
\affiliation{IMT School for Advanced Studies, P.zza San Francesco 19, 55100 Lucca (Italy)}
\author{Francesca Giuffrida}
\affiliation{Dipartimento di Chimica e Fisica, Universit\`a di Palermo, V.le delle
Scienze, Edificio 18, 90128 Palermo (Italy)}
\affiliation{IMT School for Advanced Studies, P.zza San Francesco 19, 55100 Lucca (Italy)}
\affiliation{Lorentz Institute for Theoretical Physics, University of Leiden, Niels Bohrweg 2, 2333 CA Leiden (The Netherlands)}
\author{Diego Garlaschelli}
\affiliation{IMT School for Advanced Studies, P.zza San Francesco 19, 55100 Lucca (Italy)}
\affiliation{Lorentz Institute for Theoretical Physics, University of Leiden, Niels Bohrweg 2, 2333 CA Leiden (The Netherlands)}
\affiliation{INdAM-GNAMPA Istituto Nazionale di Alta Matematica `Francesco Severi', P.le Aldo Moro 5, 00185 Rome (Italy)}
\author{Tiziano Squartini}
\affiliation{IMT School for Advanced Studies, P.zza San Francesco 19, 55100 Lucca (Italy)}
\affiliation{Scuola Normale Superiore, P.zza dei Cavalieri 7, 56126 Pisa (Italy)}
\affiliation{INdAM-GNAMPA Istituto Nazionale di Alta Matematica `Francesco Severi', P.le Aldo Moro 5, 00185 Rome (Italy)}

\date{\today}

\begin{abstract}
The study of probabilistic models for the analysis of complex networks represents a flourishing research field. Among the former, Exponential Random Graphs (ERGs) have gained increasing attention over the years. So far, only linear ERGs have been extensively employed to gain insight into the structural organization of real-world complex networks. None, however, is capable of accounting for the variance of the empirical degree distribution. To this aim, non-linear ERGs must be considered. After showing that the usual mean-field approximation forces the degree-corrected version of the two-star model to degenerate, we define a fitness-induced variant of it. Such a `softened' model is capable of reproducing the sample variance, while retaining the explanatory power of its linear counterpart, within a purely canonical framework.
\end{abstract}


\maketitle

\section{Introduction}

Network theory is employed to address problems of scientific and societal relevance, from the prediction of epidemic spreading to the identification of early-warning signals of upcoming financial crises~\cite{Colizza2006,Barrat2008,Newman2010,Pastor2015,squartini2013early,battiston2016complexity,bardoscia2017pathways,macchiati2025spectral}. As any dynamical process is strongly affected by the topology of the underlying network, one needs to individuate which higher-order properties can be traced back to lower-order ones and which, instead, are due to additional factors: this goal can be achieved by constructing ensembles of graphs whose defining properties are the same as in the real-world but the topology is random under any other respect~\cite{jaynes1957information,park2004statistical,garlaschelli2008maximum,squartini2011analytical,saracco2015randomizing,squartini2017maximum,cimini2019statistical}.

A class of models that has gained increasing attention over the years is that of Exponential Random Graphs (ERGs)~\cite{park2004statistical,Fronczak2006,Bianconi2007,garlaschelli2008maximum,squartini2011analytical,Fronczak2012,saracco2015randomizing,squartini2017maximum,cimini2019statistical}. ERGs belong to the category of canonical approaches, being induced by constraints that are \emph{soft}, i.e., constraints that can be violated by individual configurations even if their ensemble average matches the enforced value exactly. A key advantage of canonical approaches is that the expected value of topological properties can be often expressed analytically in terms of the constraints, thereby avoiding the computational cost of generating randomized networks~\cite{UnbiasedSamplingNetworkEnsembles2015}. Microcanonical approaches, instead, artificially generate many randomized variants of the observed network, enforcing constraints that are \emph{hard}, i.e., constraints that are met exactly by each graph in the ensemble~\cite{MS1995,Artzy2005,DelGenio2010,Blitzstein2011,Kim2012}; this strong requirement, however, comes at the price of non-ergodicity, high computational demand, and poor generalisability~\cite{Roberts2012}.

So far, only linear ERGs have been extensively employed to gain insight into the structural organization of real-world systems: for simple graphs, the most important one is the Undirected Binary Configuration Model (UBCM), inducing an ensemble of configurations specified by the degree sequence; a useful fitness-induced variant of it, to be employed in presence of partial information is, instead, the density-corrected Gravity Model (dcGM), solely enforcing (a proxy of) the link density while relying on node-specific quantities identified with the node strengths $\{s_i\}_{i=1}^N$, defined as $s_i=\sum_{j(\neq i)}w_{ij}$, $\forall\:i$ and representing the total volume of interactions involving each node. This choice is motivated by the evidence that in economic and financial applications a node strength is observable from balance-sheets, transactional or input-output records, while the number of its counterparties (i.e. its degree) is typically not disclosed~\cite{ReconstructionMethods2018,CimiModel2015,EnhancedGravityModelTrade2019,MethodologyEstimatingDutchInterfirmTradeNetwork2019,FunctionalStructureProductionNetworks2021,ReconstructingFirmLevelInteractionsDutchInputOutputNetwork2022}: treating strengths as exogenous fitnesses, thus, allows one to account for the heterogeneity of nodes without imposing any local constraint~\cite{UnbiasedSamplingNetworkEnsembles2015}: in formulas,

\begin{equation}
p_{ij}^\text{dcGM}=\frac{zs_is_j}{1+zs_is_j},
\end{equation}
where $z$ can be determined by requiring $L=\sum_i\sum_{j(\neq i)}p_{ij}^\text{dcGM}=\langle L\rangle$~\cite{CimiModel2015,cimini2021reconstructing}.

\begin{figure*}[t!]
\centering
\begin{minipage}{0.49\linewidth}
\centering
\includegraphics[width=\linewidth]{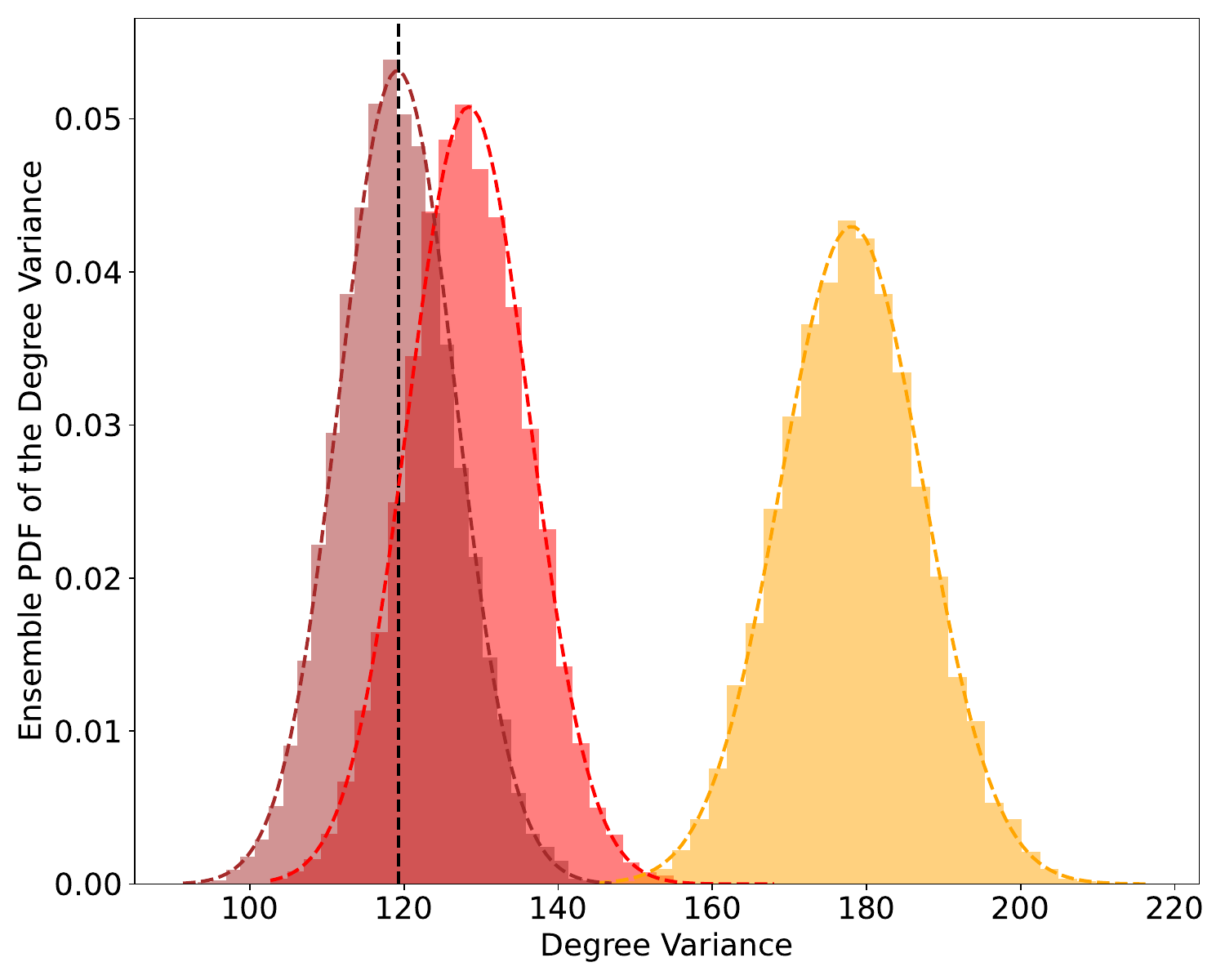}
\par\textbf{(a)}
\end{minipage}
\hfill
\begin{minipage}{0.49\linewidth}
\centering
\includegraphics[width=\linewidth]{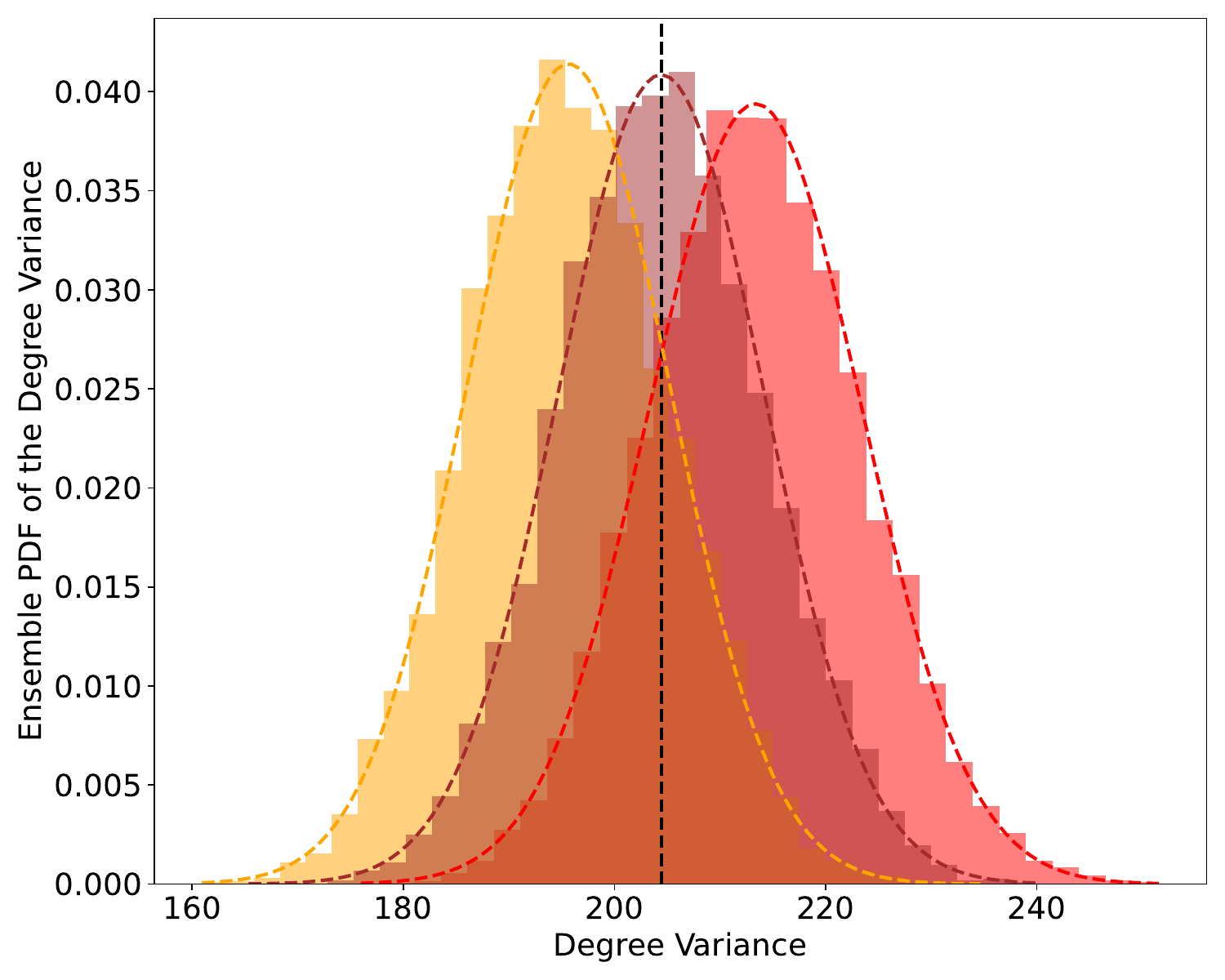}
\par\textbf{(b)}
\end{minipage}
\caption{\textbf{Reproducing the degree variance.} Analysis of eMID during the 31st week (a) and the 42nd week (b) of the year 2004. Graphical representation of the agreement between the sample variance $\text{Var}[\bm{k}]$ (black, dashed, vertical line) and its expected value $\langle\text{Var}[\bm{k}]\rangle$ under the UBCM (red), the dcGM (yellow), and the fit2SM (brown): while the UBCM steadily overestimates it and the dcGM either overestimates or underestimates it, the fit2SM correctly reproduces it. Each ensemble distribution is well approximated by a Gaussian whose parameters match the corresponding average and standard deviation.}
\label{fig:1}
\end{figure*}

Although powerful enough to reproduce many quantities of interest as accurately as the UBCM, the dcGM fails in reproducing the variance of the empirical degree distribution, either overestimating or underestimating it (see fig.~\ref{fig:1}): in the first case, larger degrees are over-estimated while smaller degrees are under-estimated; in the second one, larger degrees are under-estimated while smaller degrees are over-estimated. Accurately reproducing the variance of the degree distribution is crucial for a variety of applications, such as correctly assessing the systemic risk related to the interconnections of a given financial system~\cite{battiston2012debtrank,bardoscia2017pathways} or the epidemic threshold associated with the interlinkages of a given social system - reading $\langle k\rangle/\langle k^2\rangle$, hence being governed by the first two moments of the degree distribution~\cite{pastor2001epidemic,Pastor2015}; another example is provided by the expected consensus time of the so-called coalescing random walks, whose estimation requires calculating the first two moments of the degree distribution as well~\cite{SoodRedner2005,SoodAntalRedner2008}.

The aim of this work is to investigate whether it is possible to define a minimal model capable of accurately reproducing both the first and second moments of empirical degree distributions, without resorting to microcanonical constraints. To this end, we consider the class of ERGs with non-linear constraints and introduce a fitness-induced variant of the two-star model that can be efficiently and reliably implemented. We test it on the transaction-level data constituting the overnight segment of the Electronic Market for Interbank Deposits (eMID), a screen-based market for unsecured deposits~\cite{macchiati2025spectral,IoriOvernightMoneyMarket2008,FingerFrickeLuxEMID2012}. Although the raw records are directed and weighted (lender, borrower, notional amount), here, we symmetrize and binarize exposures within each, considered time window, i.e. `daily', `weekly', `monthly', `quarterly' and `yearly' (the description of each aggregation procedure is reported in Appendix~\hyperlink{AppA}{A}).

\section{The simplest non-linear constraint}

The simplest non-linear constraint is represented by the total number of two-stars~\cite{park2004statistical,park2004solution}, i.e., 

\begin{equation}
S=\sum_i\sum_{j(>i)}V_{ij}=\sum_i\sum_{j(>i)}\sum_ma_{im}a_{jm};
\end{equation}
since

\begin{align}
S&=\frac{1}{2}\sum_m\sum_i\sum_{j(\neq i)}a_{im}a_{jm}\nonumber\\
&=\frac{1}{2}\sum_m\sum_ia_{im}\left(\sum_ja_{jm}-a_{im}\right)\nonumber\\
&=\frac{1}{2}\sum_m\left(\sum_ia_{im}k_m-\sum_ia_{im}\right)\nonumber\\
&=\frac{1}{2}\sum_m(k_m^2-k_m)\nonumber\\
&=\frac{1}{2}\sum_mk_m^2-L,
\end{align}
inverting such a relationship leads the second moment of the empirical degree distribution to be re-writable as

\begin{align}
\overline{k^2}=\frac{\sum_mk_m^2}{N}=\frac{2S}{N}+\frac{2L}{N}
\end{align}
and its variance (hereby, sample variance) as

\begin{align}
\text{Var}[\bm{k}]=\overline{k^2}-\overline{k}^2
&=\frac{2S}{N}+\frac{2L}{N}\left(1-\frac{2L}{N}\right).
\end{align}

As a consequence, its expected value reads

\begin{align}\label{eq:4}
\langle\text{Var}[\bm{k}]\rangle&=\frac{2\langle S\rangle}{N}+\frac{2\langle L\rangle}{N}-\frac{4\langle L^2\rangle}{N^2}\nonumber\\
&=\frac{2\langle S\rangle}{N}+\frac{2\langle L\rangle}{N}
-\frac{4\text{Var}[L]}{N^2}-\frac{4\langle L\rangle^2}{N^2}\nonumber\\
&=\frac{2\langle S\rangle}{N}-\frac{4\text{Var}[L]}{N^2}+\frac{2\langle L\rangle}{N}\left(1-\frac{2\langle L\rangle}{N}\right),
\end{align}
an expression showing that reproducing the sample variance requires (at least) the total number of links to be reproduced \emph{exactly}.

\section{A microcanonical approach?}

Since the result above suggests the microcanonical road as the only viable one, a natural choice would be that of considering the microcanonical version of the two-star model (m2SM), constraining both $L$ and $S$ exactly: since $\langle L\rangle=L$, $\text{Var}[L]=0$, and $\langle S\rangle=S$, one would obtain $\langle\text{Var}[\bm{k}]\rangle=\text{Var}[\bm{k}]$. This model is, however, \emph{homogeneous} and, therefore, unable to reproduce the degree sequence.

We would be, thus, tempted to refine it by constraining \emph{both} the total number of two-stars \emph{and} the degree sequence. Yet, since

\begin{align}\label{eq:5}
S-\langle S\rangle&=\frac{1}{2}\left[\sum_i(k_i^2-k_i)-\sum_i(\langle k_i^2\rangle-\langle k_i\rangle)\right]\nonumber\\
&=\frac{1}{2}\sum_i(k_i^2-\langle k_i^2\rangle)\nonumber\\
&=\frac{1}{2}\sum_i(\langle k_i\rangle^2-\langle k_i^2\rangle)\nonumber\\
&=-\frac{1}{2}\sum_i\text{Var}[k_i],
\end{align}
satisfying both (sets of) constraints microcanonically is equivalent to requiring that $\text{Var}[k_i]=0$, $\forall\:i$. In words, our results indicate that the microcanonical version of the Configuration Model (mCM) is the (simplest) one guaranteeing $\langle k_i\rangle=k_i$, $\forall\:i$ and $\langle S\rangle=S$. Moreover, as the relationship

\begin{equation}
\text{Var}[L]=\text{Var}\left[\frac{\sum_ik_i}{2}\right]=\frac{\sum_i\text{Var}[k_i]+2\sum_{l<n}\text{Cov}[k_l,k_n]}{4}
\end{equation}
leads to the expression

\begin{align}
\langle\text{Var}[\bm{k}]\rangle=&\:\frac{2\langle S\rangle}{N}-\frac{\sum_i\text{Var}[k_i]+2\sum_{l<n}\text{Cov}[k_l,k_n]}{N^2}\nonumber\\
&+\frac{2\langle L\rangle}{N}\left(1-\frac{2\langle L\rangle}{N}\right),
\end{align}
the mCM would also guarantee that $\langle\text{Var}[\bm{k}]\rangle=\text{Var}[\bm{k}]$: in words, constraining the entire degree sequence \emph{exactly} would lead to reproduce the sample variance as well.

Although the (microcanonical) requirements are clear, the way to realize them is much less so; moreover, the problems affecting the microcanonical approaches mentioned in the introductory paragraph lead us to discard this route. Is a canonical way out viable?

\section{The mean-field approximation}

Handling non-linear constraints within a canonical framework is usually achieved by adopting the mean-field approximation (see also Appendix~\hyperlink{AppB}{B})~\cite{park2004statistical,park2004solution}. 

Constraining both the number of two-stars and the degree sequence within such a framework leads to the \emph{degree-corrected Two-Star Model} (dc2SM): formally introduced in Appendix~\hyperlink{AppC}{C}, its generic probability coefficient reads

\begin{equation}\label{eq:7}
p_{ij}^\text{dc2SM}=\frac{e^{-(\alpha_i+\alpha_j)-\beta(k_i+k_j)}}{1+e^{-(\alpha_i+\alpha_j)-\beta(k_i+k_j)}}=\frac{x_ix_jy^{k_i+k_j}}{1+x_ix_jy^{k_i+k_j}};
\end{equation}
the dc2SM is, however, unable to match both (sets of) constraints, as reproducing the empirical number of two-stars implies underestimating the degree of at least one node (see also Appendix~\hyperlink{AppC}{C}). The explanation of such a behavior lies in a simple observation: as the mean-field approximation scheme requires that

\begin{equation}
\text{Var}[k_i]=\sum_{j(\neq i)}p_{ij}(1-p_{ij}),\quad\forall\:i,
\end{equation}
letting eq.~\ref{eq:5} vanish requires the generic probability coefficient of a canonical model to degenerate, becoming either 0 or 1. In other words, constraining both the degrees and the number of two-stars within the mean-field approximation scheme forces the model to become deterministic, individuating the observed configuration\footnote{The asterisk indicates (the value of the quantities measured on) the observed configuration $\mathbf{A}^*$.} $\mathbf{A}^*$ as the only admissible one.

\begin{figure*}[t!]
\centering
\begin{minipage}{0.32\linewidth}
\centering
\includegraphics[width=\linewidth]{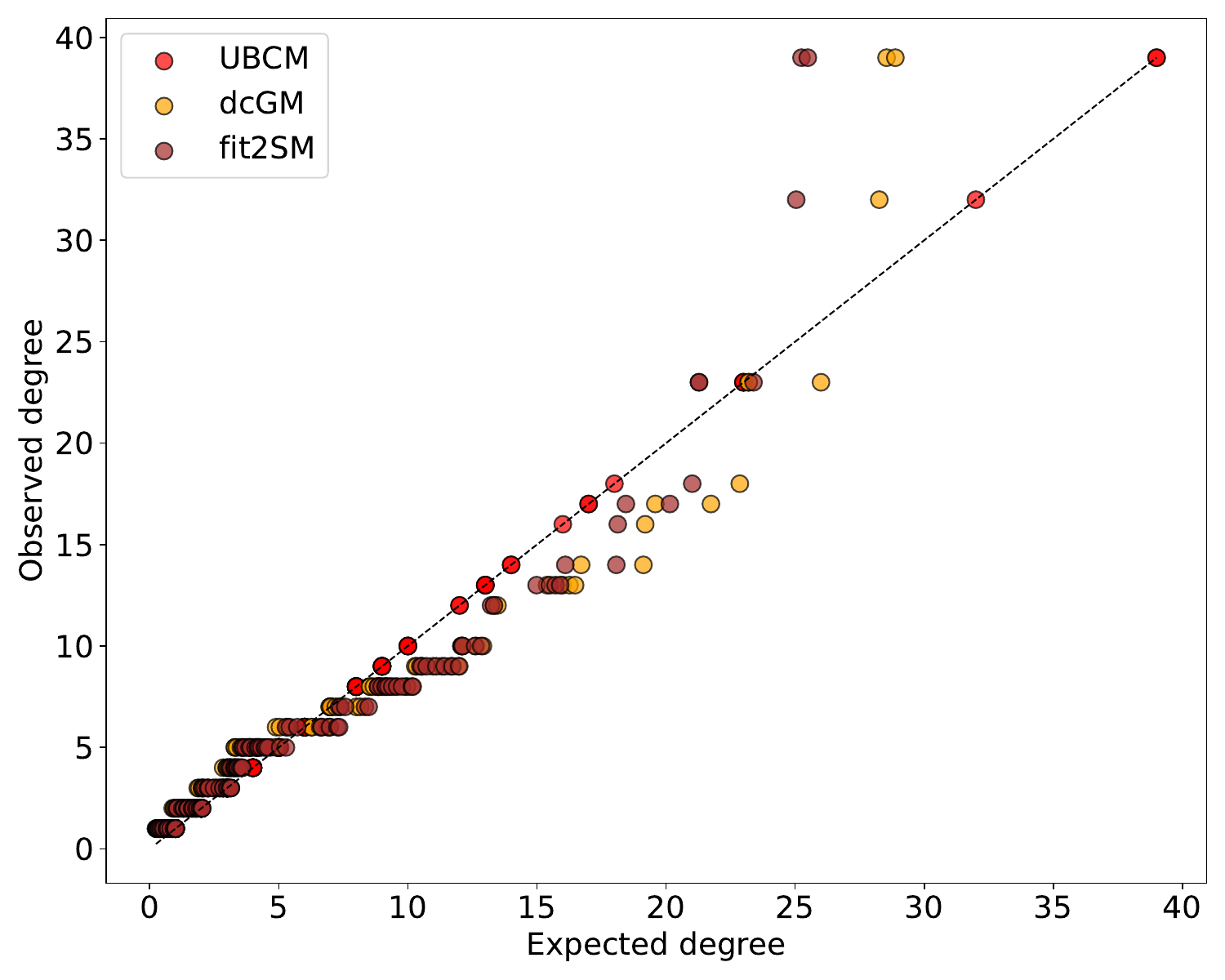}
\par\textbf{(a)}
\end{minipage}
\hfill
\begin{minipage}{0.32\linewidth}
\centering
\includegraphics[width=\linewidth]{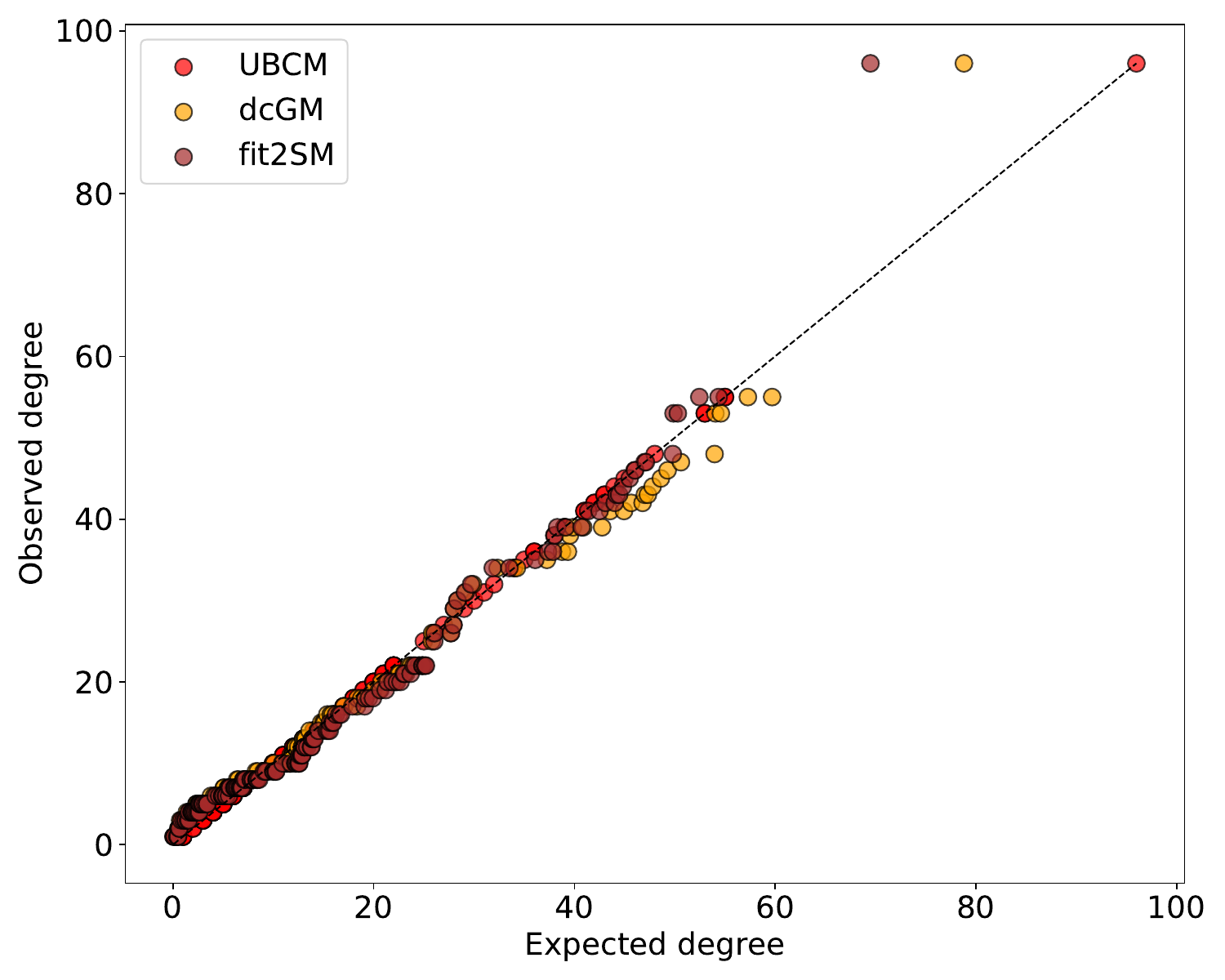}
\par\textbf{(b)}
\end{minipage}
\hfill
\begin{minipage}{0.32\linewidth}
\centering
\includegraphics[width=\linewidth]{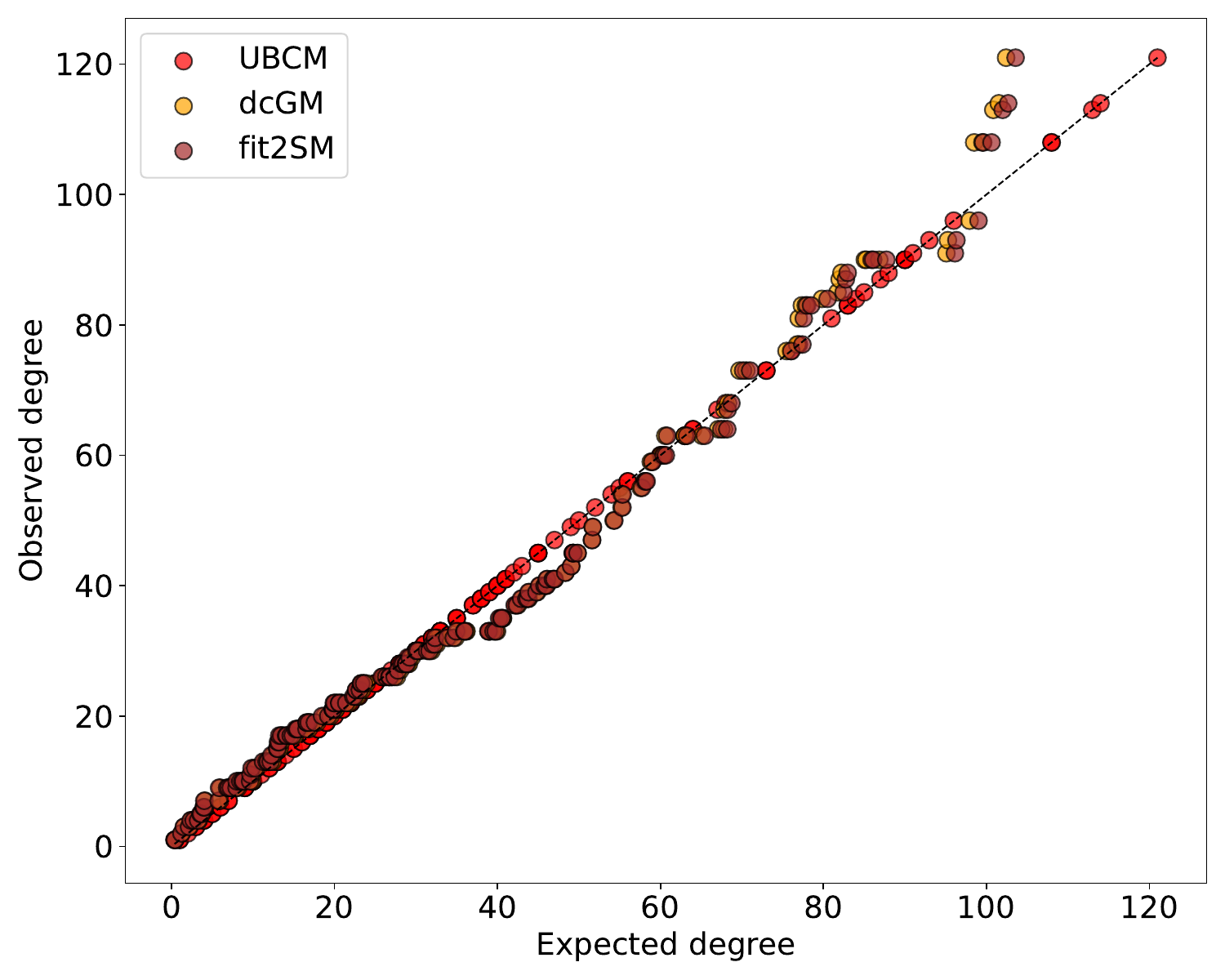}
\par\textbf{(c)}
\end{minipage}
\par\vspace{0.4em}
\begin{minipage}{0.32\linewidth}
\centering
\includegraphics[width=\linewidth]{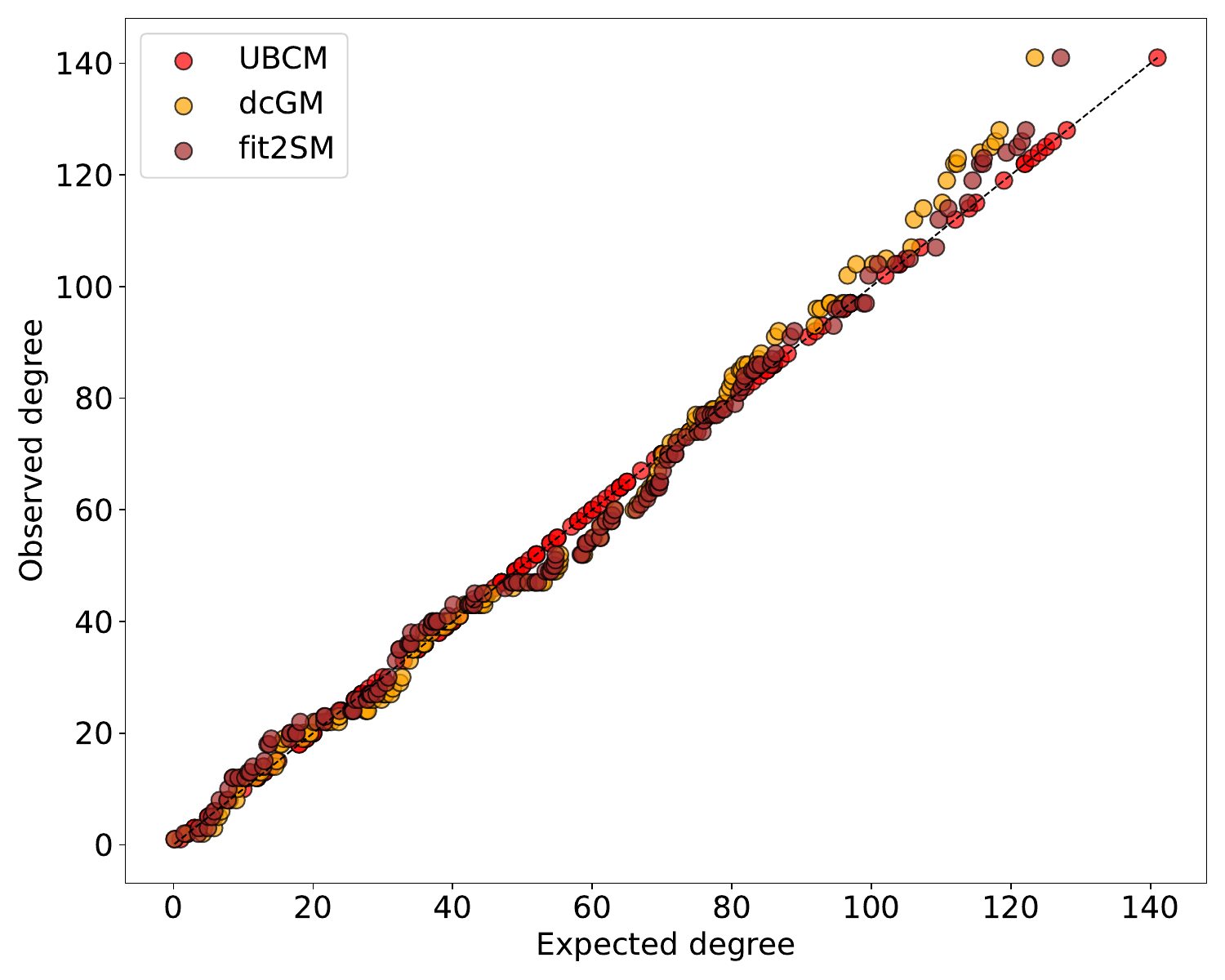}
\par\textbf{(d)}
\end{minipage}
\hspace{0.02\linewidth}
\begin{minipage}{0.32\linewidth}
\centering
\includegraphics[width=\linewidth]{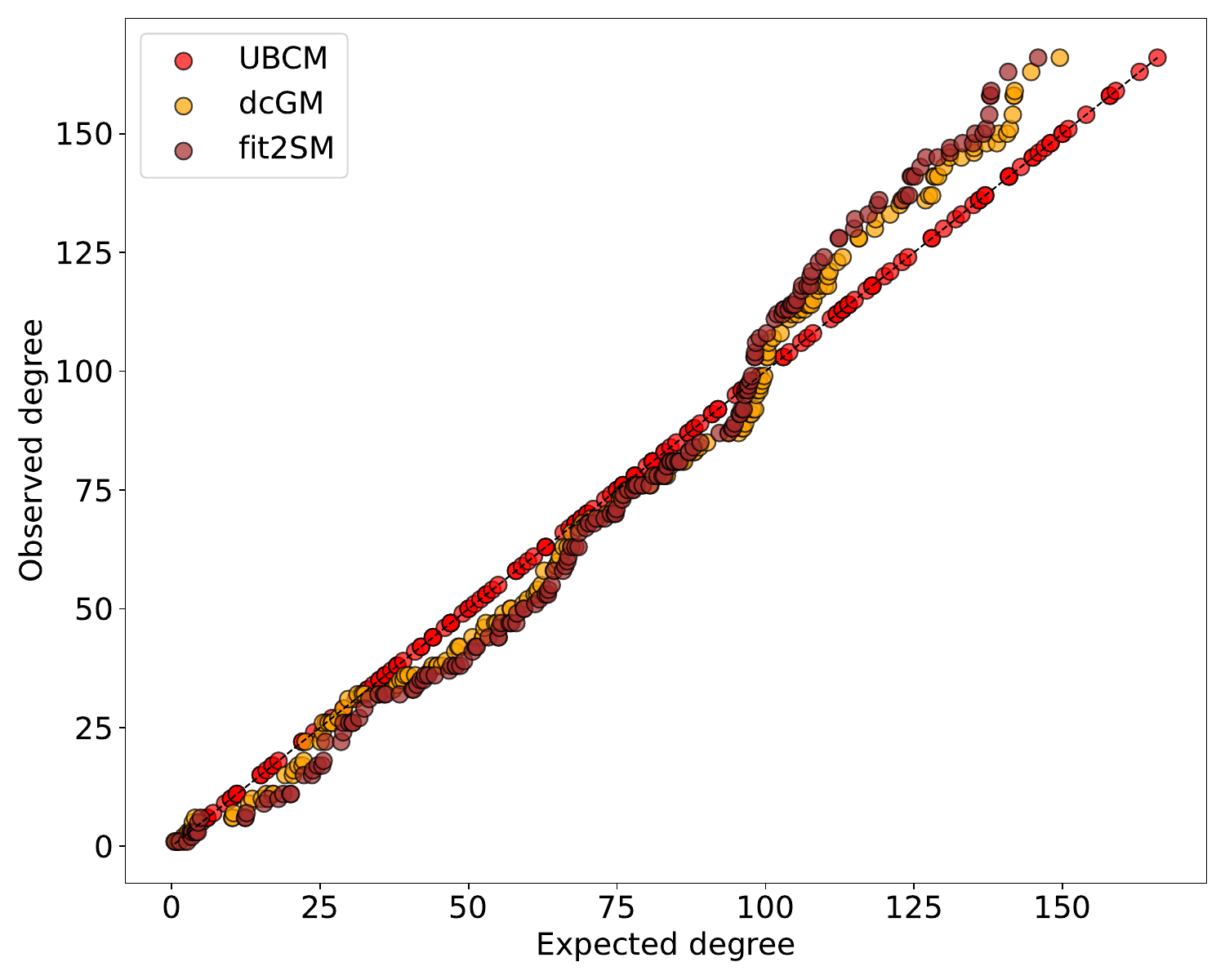}
\par\textbf{(e)}
\end{minipage}
\caption{\textbf{Reproducing the degrees across temporal aggregations.} Quantile-quantile plots comparing the observed degrees and the expected ones under the UBCM (red), the dcGM (yellow) and the fit2SM (brown), across five aggregations of the eMID interbank market (panels~(a--e): daily, weekly, monthly, quarterly and yearly). These snapshots are selected by drawing one calendar year at random (here, $2001$) and, conditionally on it, drawing at random one quarter (here, Q3), one month (here, April), one ISO week (here, week 2) and one trading day (here, $2001$-$11$-$21$). As it can be appreciated, the dcGM and the fit2SM perform very similarly, and both satisfactorily, in reproducing the degree sequence induced by the corresponding snapshot.}
\label{fig:2}
\end{figure*}

\section{A canonical way out}

Let us now ask ourselves if an alternative, canonical road exists. To this aim, let us, first, notice that $\text{Var}[L]$ is divided by $N^2$ in eq.~\ref{eq:4}: such an addendum may, thus, be expected not to play a relevant role. As a consequence, we are allowed to consider the expression:

\begin{align}
\langle\text{Var}[\bm{k}]\rangle&\simeq\frac{2\langle S\rangle}{N}+\frac{2\langle L\rangle}{N}\left(1-\frac{2\langle L\rangle}{N}\right),
\end{align}
requiring the total number of links and the total number of two-stars to be reproduced \emph{on average}. Since any linear ERG reproduces the total number of links, the failure in reproducing the sample variance boils down to the failure in reproducing the total number of two-stars: an overestimation/underestimation of the latter leads, in fact, to an overestimation/underestimation of the former - this is precisely the case of the canonical version of the Configuration Model (UBCM), that overestimates the total number of two-stars (see eq.~\ref{eq:5}), hence predicting $\langle\text{Var}[\bm{k}]\rangle\geq\text{Var}[\bm{k}]$ (see fig.~\ref{fig:1}).

At this point, a natural choice would be that of considering the canonical version of the Two-Star Model (2SM): within the mean-field approximation scheme, however, the 2SM is defined by the position

\begin{equation}
p=\frac{xy^{k_i+k_j}}{1+xy^{k_i+k_j}}=\frac{xy^{2(N-1)p}}{1+xy^{2(N-1)p}}
\end{equation}
which, again, makes it a homogeneous model\footnote{Limiting ourselves to the position $p_{ij}=xy^{k_i+k_j}/(1+xy^{k_i+k_j})$ would, in fact, lead to an inconsistency since the information on the degrees is not supposed to be available.} (see also Appendix~\hyperlink{AppC}{C})~\cite{park2004statistical,park2004solution}.

\begin{figure*}[t!]
\centering
\begin{minipage}{0.49\linewidth}
\centering
\includegraphics[width=\linewidth]{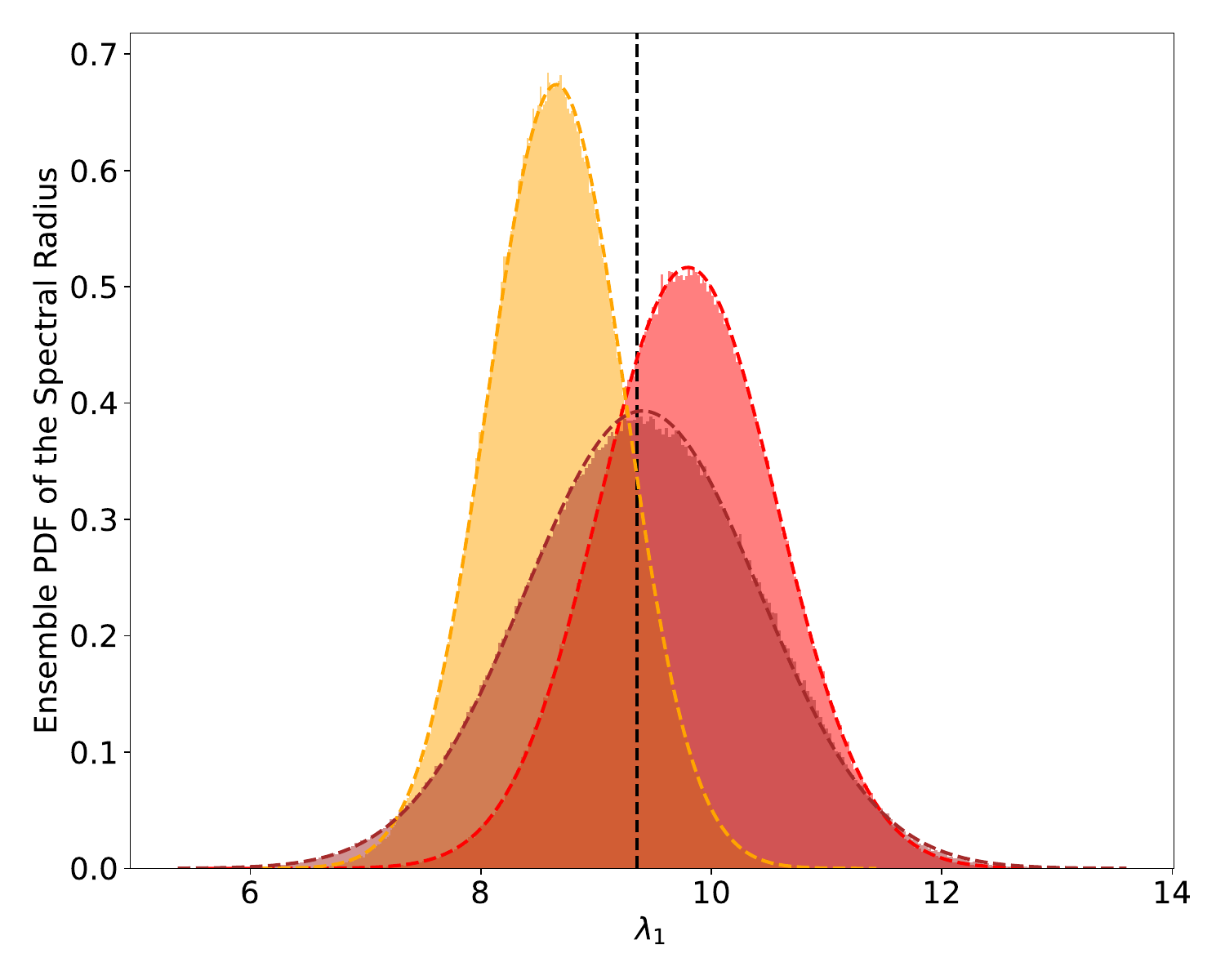}
\par\textbf{(a)}
\end{minipage}
\hfill
\begin{minipage}{0.483\linewidth}
\centering
\includegraphics[width=\linewidth]{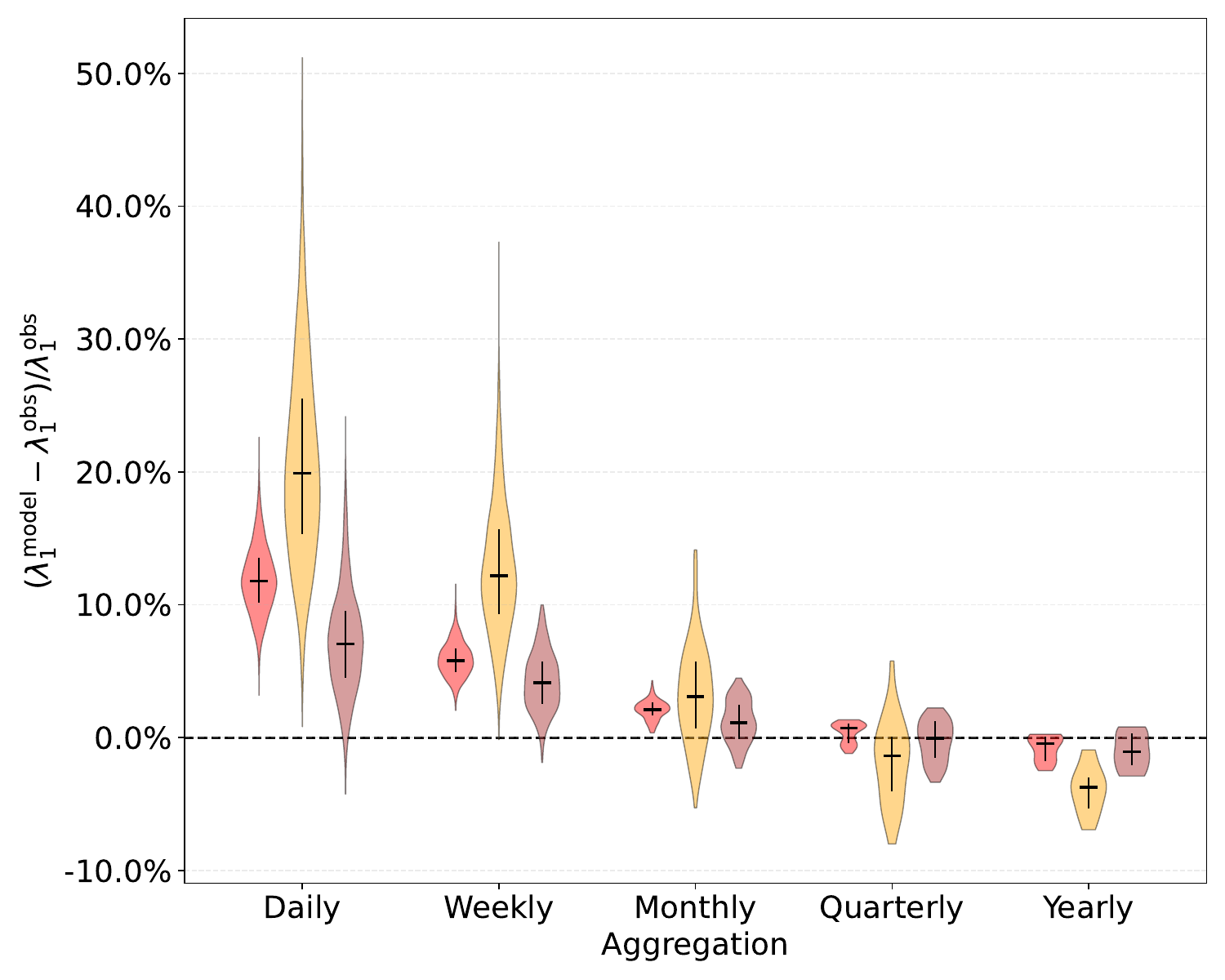}
\par\textbf{(b)}
\end{minipage}
\caption{\textbf{Estimating the spectral radius of real-world configurations.} Panel~(a): distributions of the spectral radius induced by $M=10^3$ configurations sampled from the Chung-Lu model defined by the topology of the eMID snapshot corresponding to the day $2010$-$07$-$19$, according to the UBCM (red), the dcGM (yellow), and the fit2SM (brown). The three models were solved for each of the $M=10^3$, generated configurations and the corresponding ensembles explicitly sampled in order to obtain an estimation of $\pi_1$: what we find is that $\lambda_1^\text{dcGM}<\pi_1\lesssim\lambda_1^\text{fit2SM}<\lambda_1^\text{UBCM}$. In words, while the UBCM overestimates the spectral radius of the generative model (black, dashed, vertical line) and the dcGM underestimates it, the fit2SM correctly reproduces it. Each ensemble distribution is well approximated by a Gaussian whose parameters match the corresponding average and standard deviation. Panel~(b): violin plots (summarizing the distributions) of the relative error in reproducing the empirical spectral radius $\lambda_1^{\text{obs}}$, across snapshots and temporal aggregations. The fit2SM (brown) yields smaller errors than \emph{i)} the dcGM (yellow) at all time-scales and \emph{ii)} the UBCM (red) at the daily and weekly time-scales.}
\label{fig:3}
\end{figure*}

Constructing a fitness-induced variant of the 2SM (fit2SM) is, nevertheless, feasible: to this aim, it is enough to replace $x_i$ with $\sqrt{z}s_i$\footnote{To ease numerical manipulations, we have rescaled the strengths, dividing them by their arithmetic mean.} and $k_i$ with its expectation under the same model in eq.~\ref{eq:7}. These substitutions lead to the generic probability coefficient reading

\begin{equation}\label{eq:score_fitTwoStar}
p_{ij}^\text{fit2SM}=\frac{zs_is_j y^{\kappa_i+\kappa_j}}{1+zs_is_j y^{\kappa_i+\kappa_j}}
\end{equation}
where $\langle k_i\rangle_\text{fit2SM}\equiv\kappa_i$. The two, global parameters $z$ and $y$ can be estimated by solving the (system constituted by the) two non-linear coupled equations reading

\begin{equation}
\left\{
\begin{aligned}
L^*&=\sum_i\sum_{j(>i)}p_{ij}^\text{fit2SM}=\langle L\rangle\\
S^*&=\sum_i\sum_{j(>i)}\sum_mp_{im}^\text{fit2SM}p_{jm}^\text{fit2SM}=\langle S\rangle
\end{aligned}
\right.
\end{equation}
and re-writable~\cite{vallarano2021fast} as

\begin{equation}
\left\{
\begin{aligned}
&z_{(t+1)}=\frac{L^*}{\sum_i\sum_{j(>i)}\frac{s_i^*s_j^*y_{(t)}^{\kappa_i+\kappa_j}}{1+z_{(t)}s_i^*s_j^*y_{(t)}^{\kappa_i+\kappa_j}}}\\
&y_{(t+1)}\\
&=\frac{S^*}{\sum_i\sum_{j(>i)}\sum_m\frac{z_{(t)}^2s_i^*(s_m^*)^2s_j^*y_{(t)}^{\kappa_i+2\kappa_m+\kappa_j-1}}{\left(1+z_{(t)}s_i^*s_m^*y_{(t)}^{\kappa_i+\kappa_m}\right)\left(1+z_{(t)}s_j^*s_m^*y_{(t)}^{\kappa_j+\kappa_m}\right)}}
\end{aligned}
\right.
\end{equation}

Notice that a third set of consistency equations concerning the degrees should be added. Operatively, we should \emph{i)} initialise the degrees in some way; \emph{ii)} solve the system above, obtaining $z$ and $y$ at the `epoch' 1 (i.e. $z_1$ and $y_1$); \emph{iii)} repeat the steps \emph{i)} and \emph{ii)} to calculate the corresponding values of the degrees and use them to obtain $z$ and $y$ at the subsequent `epochs'. In formulas

\begin{equation}
\kappa_i^{(e+1)}=\sum_{j(\neq i)}\frac{z_{(e)}s_i^*s_j^*y_{(e)}^{\kappa_i^{(e)}+\kappa_j^{(e)}}}{1+z_{(e)}s_i^*s_j^*y_{(e)}^{\kappa_i^{(e)}+\kappa_j^{(e)}}},\quad\forall\:i
\end{equation}
with $e$ indicating the `epoch' of the resolution. As we show in Appendix~\hyperlink{AppD}{D}, however, stopping at the `epoch' 0, by posing

\begin{equation}
\kappa_i^{(0)}=\sum_{j(\neq i)}p_{ij}^\text{dcGM},\quad\forall\:i,
\end{equation}
is (already) enough to achieve good results in terms of speed and accuracy of resolution. The system above ensures that the fit2SM correctly reproduces the total number of links, the total number of two-stars and the sample variance, besides accounting for the nodes heterogeneity.

\begin{figure*}[t!]
\centering
\includegraphics[width=\linewidth]{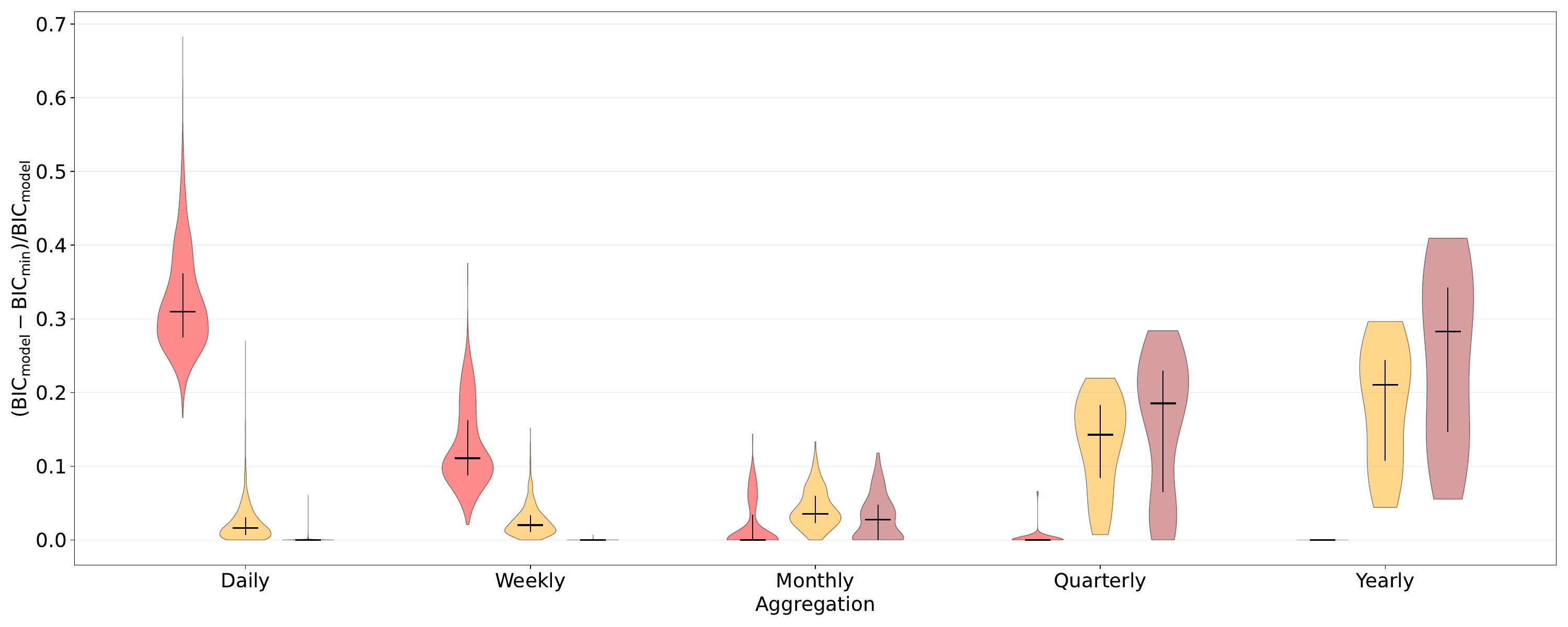}
\caption{\textbf{Model comparison across temporal aggregations.} For each snapshot and aggregation level, we show the violin plots (summarising the distributions) of the relative error $\delta\text{BIC}_{\text{model}}$, defined in eq.~\ref{eq:bic}. Upon remembering that a null value of such an index identifies the best-performing model on the considered snapshot, let us notice that the score induced by the fit2SM (brown) concentrates near zero - hence such a model dominates over the competing ones - at finer aggregations, while the UBCM (red) is the winner at coarser aggregations.}
\label{fig:4}
\end{figure*}

\section{Testing the performance\\of the fit2SM}

\subsection{Reproducing a network's local properties}

Since the degree sequence is a standard diagnostic in network reconstruction exercises, we start our analysis by testing the accuracy of the fit2SM in reproducing it. We do so by means of the quantile-quantile (QQ) plot: given the empirical sequence $\{k_i\}_{i=1}^N$ and a model-based, expected one $\{\langle k_i\rangle\}_{i=1}^N$, we sort both in increasing order and scatter the corresponding pairs of values\footnote{Recall that the quantile $q$ of a set of (ordered) values indicates the percentage $q$ of smallest values. Given the empirical degree sequence and the model-based, expected one, we, thus, compute their quantile functions by sorting both in increasing order and scattering the $r$-th order statistics $\langle k\rangle_r$ versus $k_r$, $r=1,\dots,N$.}; while perfect agreement would place all points on the identity line, mis-estimations of the values would be revealed by systematic deviations.

Let us, now, select five snapshots by drawing one calendar year at random and, conditionally on it, drawing one quarter, one month, one ISO week, and one trading day at random as well. Figure~\ref{fig:2} reports the resulting QQ plots for year 2001, quarter Q3, month April, ISO week 2, and day $2001$-$11$-$21$; the visual inspection is accompanied by the accuracy indicators named \emph{average relative error}, defined as $\text{ARE}=N^{-1}\sum_i|k_i-\langle k_i\rangle|/k_i$, and \emph{maximum relative error}, defined as $\text{MRE}=\max_i\{|k_i-\langle k_i\rangle|/k_i\}$.

While the UBCM reproduces the degrees by construction, the dcGM and the fit2SM yield comparable errors, with the $\text{ARE}$ ranging from $\simeq 0.51$ to $\simeq 0.80$ and the $\text{MRE}$ from $\simeq 6.60$ to $\simeq 22.2$, across aggregations: more specifically, while the fit2SM achieves a smaller $\text{ARE}$ at the daily, monthly and quarterly scales, the dcGM performs better at the weekly and yearly scales; at the node level, the fit2SM yields a smaller ARE in the $65\%$ of cases at the daily scale, in the $61\%$ of cases at the weekly scale, in the $41\%$ of cases at the monthly scale, in the $38\%$ of cases at the quarterly scale and in the $54\%$ of cases at the yearly scale. Repeating the procedure above for other, randomly selected snapshots yields analogous results.

As these observations point out, the predictions of the degrees returned by the dcGM and the fit2SM turn out to be very close: as the fit2SM is not conceived to improve the matching of the degrees but to incorporate a non-linear constraint to reproduce their variance, the related `benefits' are expected to emerge more clearly when inspecting properties that depend explicitly on the second moment - such as the spectral radius (see below).

\begin{figure*}[t!]
\centering
\begin{minipage}{0.49\linewidth}
\centering
\includegraphics[width=\linewidth]{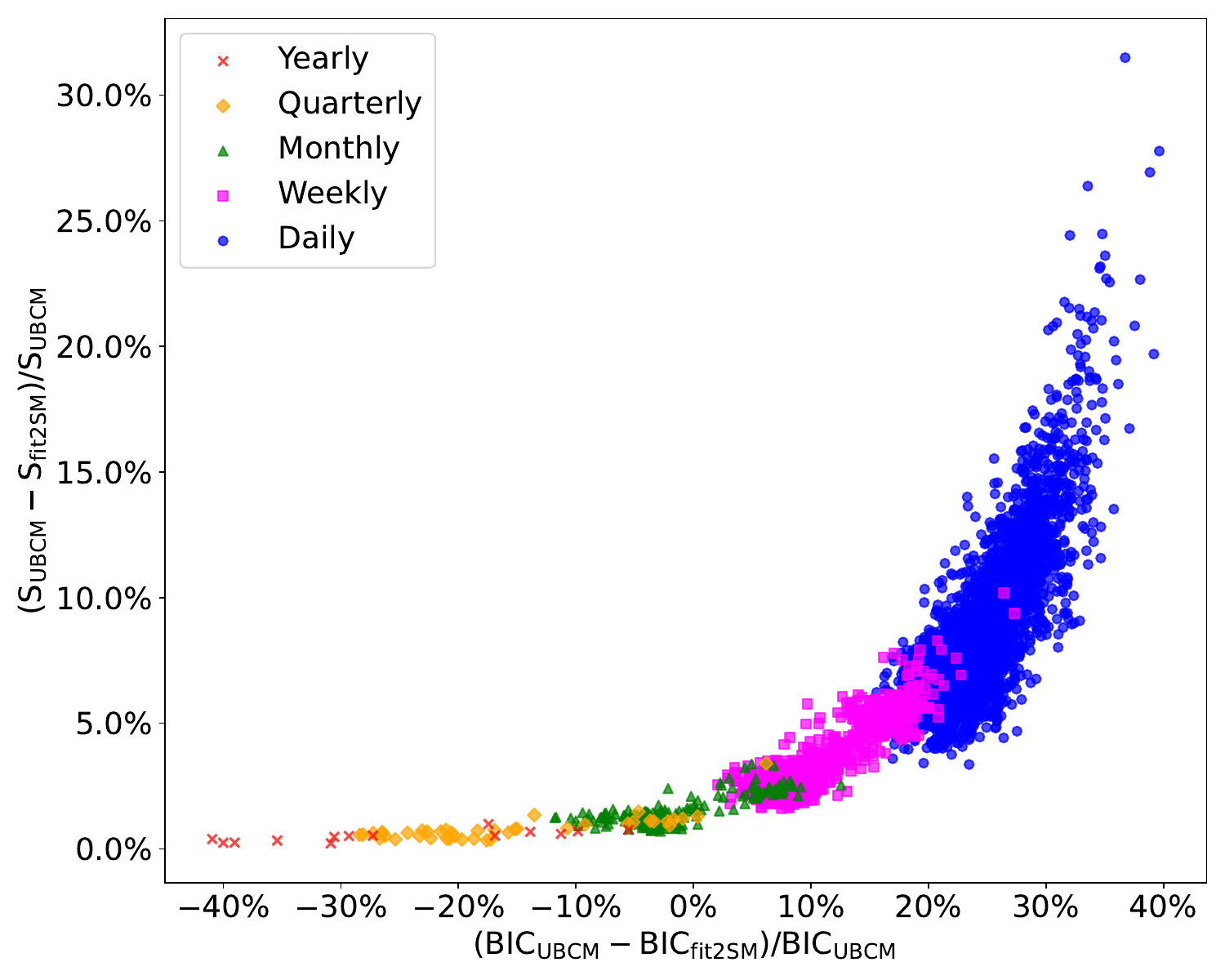}
\par\textbf{(a)}
\end{minipage}
\hfill
\begin{minipage}{0.49\linewidth}
\centering
\includegraphics[width=\linewidth]{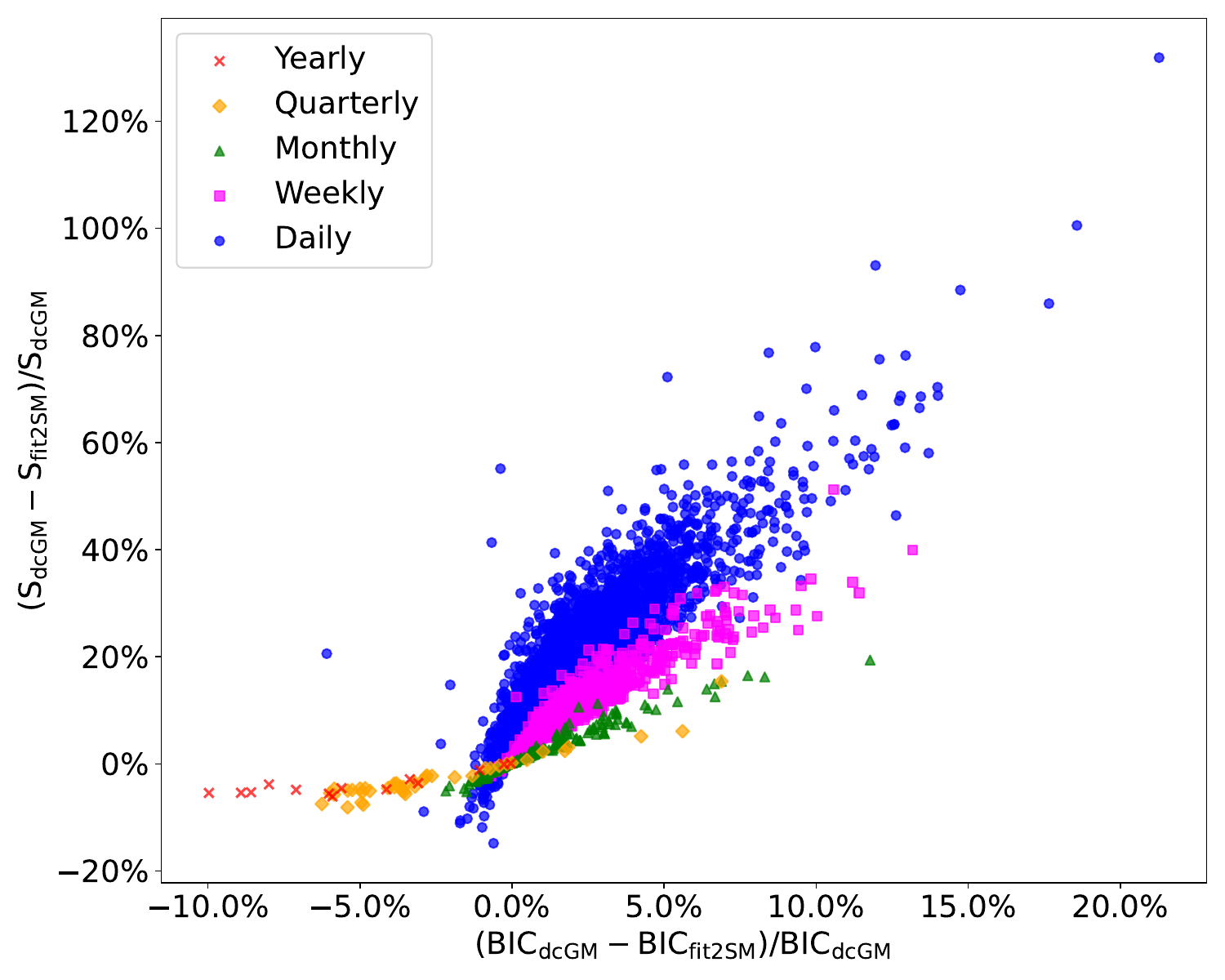}
\par\textbf{(b)}
\end{minipage}
\caption{\textbf{UBCM versus fit2SM.} Panel~(a): relative improvement of the fit2SM versus the relative error in estimating the number of two-stars with respect to the UBCM. Panel~(b): relative improvement of the fit2SM versus the relative error in estimating the number of two-stars with respect to the dcGM. A trend appear: the larger the error in estimating the total number of two-stars, the more the fit2SM outperforms the other models, an evidence suggesting that a poorer estimation of the total number of two-stars corresponds to a worse statistical fit of the eMID structure. This effect is more pronounced for sparser configurations, on which both the UBCM and the dcGM tend to commit larger errors - while the UBCM steadily overestimates $S$, the dcGM either overestimate or underestimate it. Each point is a network at a given level of aggregation.}
\label{fig:5}
\end{figure*}

\subsection{Reproducing a network spectral properties}

As a second test, let us inspect the accuracy of the fit2SM in reproducing the spectral radius, hereby indicated with $\pi_1$, of a given network. To this aim, let us follow~\cite{macchiati2025spectral} and consider the fully-controllable case represented by the Chung-Lu model~\cite{chung2002connected}. A way to identify $\pi_1$ in case $p_{ij}=k_ik_j/2L$, $\forall\:i,j$ rests upon the relationship

\begin{equation}
\mathbf{P}=\frac{\mathbf{k}\otimes\textbf{k}}{2L}=\frac{\ket{k}\bra{k}}{2L},
\end{equation}
indicating that, in such a case, the matrix $\mathbf{P}$ can be obtained as the direct product of the vector of degrees with itself. Employing the bra-ket formalism allows the calculations to be carried out quite easily, i.e., as

\begin{equation}
\mathbf{P}\ket{k}=\frac{\ket{k}\bra{k}}{2L}\ket{k}=\frac{\bra{k}\ket{k}}{2L}\ket{k}
\end{equation}
where $\bra{k}\ket{k}=\sum_ik_i^2$. Since $\mathbf{P}$ obeys the Perron-Frobenius theorem~\cite{horn1985matrix}, the equation above allows us to identify the value of its spectral radius quite straightforwardly as $\pi_1=\bra{k}\ket{k}/2L=\sum_ik_i^2/2L$. The Chung-Lu model is, however, defined by the position $p_{ij}=k_ik_j/2L$, $\forall\:i\neq j$, a piece of evidence leading us to write

\begin{equation}
\pi_1\simeq\frac{\bra{k}\ket{k}}{2L}=\sum_i\frac{k_i^2}{2L}=\frac{N}{2L}\sum_i\frac{k_i^2}{N}=\frac{\overline{k^2}}{\overline{k}};
\end{equation}
in other words, the Chung-Lu model represents a convenient benchmark to make the connection between the accuracy of a network model in reproducing the first two moments of the degree distribution and the one in reproducing the spectral properties of the corresponding configuration explicit, via the dependence of $\pi_1$ on $\overline{k}$ and $\overline{k^2}$.

Let us now consider the eMID snapshot corresponding to the day $2010$-$07$-$19$ and calibrate the model above on it - a choice allowing us to deal with a graphical degree sequence. Afterwards, let us generate $M=10^3$ configurations and treat each of them as a plausible, empirical one; solving our models on the latter leads to three distributions of $M=10^3$ estimates of the spectral radius each (hereby indicated with $\lambda_1$, to distinguish them from the reference value $\pi_1$; see also Appendix~\hyperlink{AppE}{E}): as panel~(a) of fig.~\ref{fig:3} shows, while the UBCM overestimates the spectral radius of the generative model ($\lambda_1^\text{UBCM}\simeq 9.80>\pi_1\simeq 9.35$) and the dcGM underestimates it ($\lambda_1^\text{dcGM}\simeq 8.66<\pi_1\simeq 9.35$), the fit2SM correctly reproduces it ($\lambda_1^\text{fit2SM}\simeq 9.40\gtrsim\pi_1\simeq 9.35$).

\begin{figure*}[t!]
\centering
\begin{minipage}{0.49\linewidth}
\centering
\includegraphics[width=\linewidth]{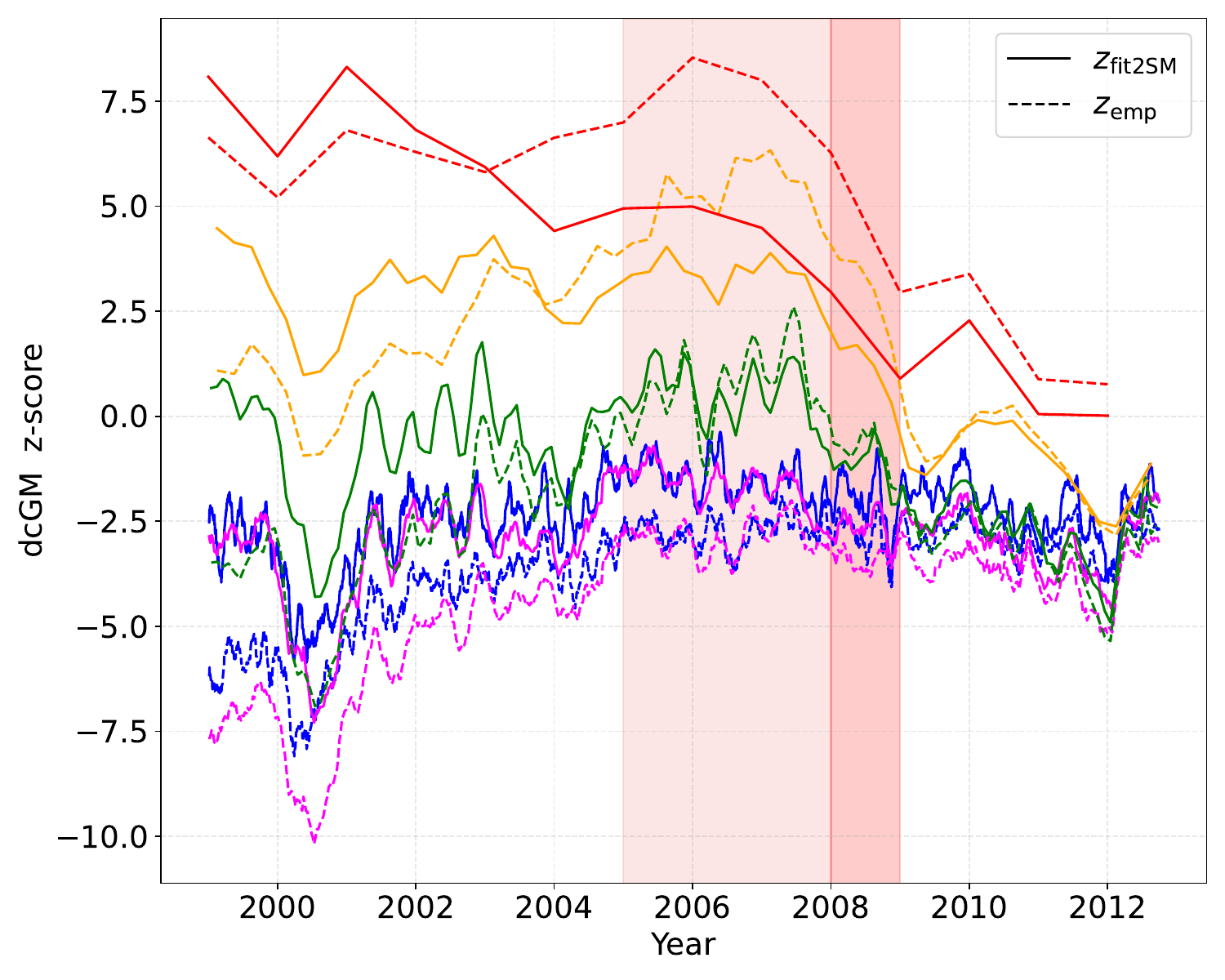}
\par\textbf{(a)}
\end{minipage}
\hfill
\begin{minipage}{0.49\linewidth}
\centering
\includegraphics[width=\linewidth]{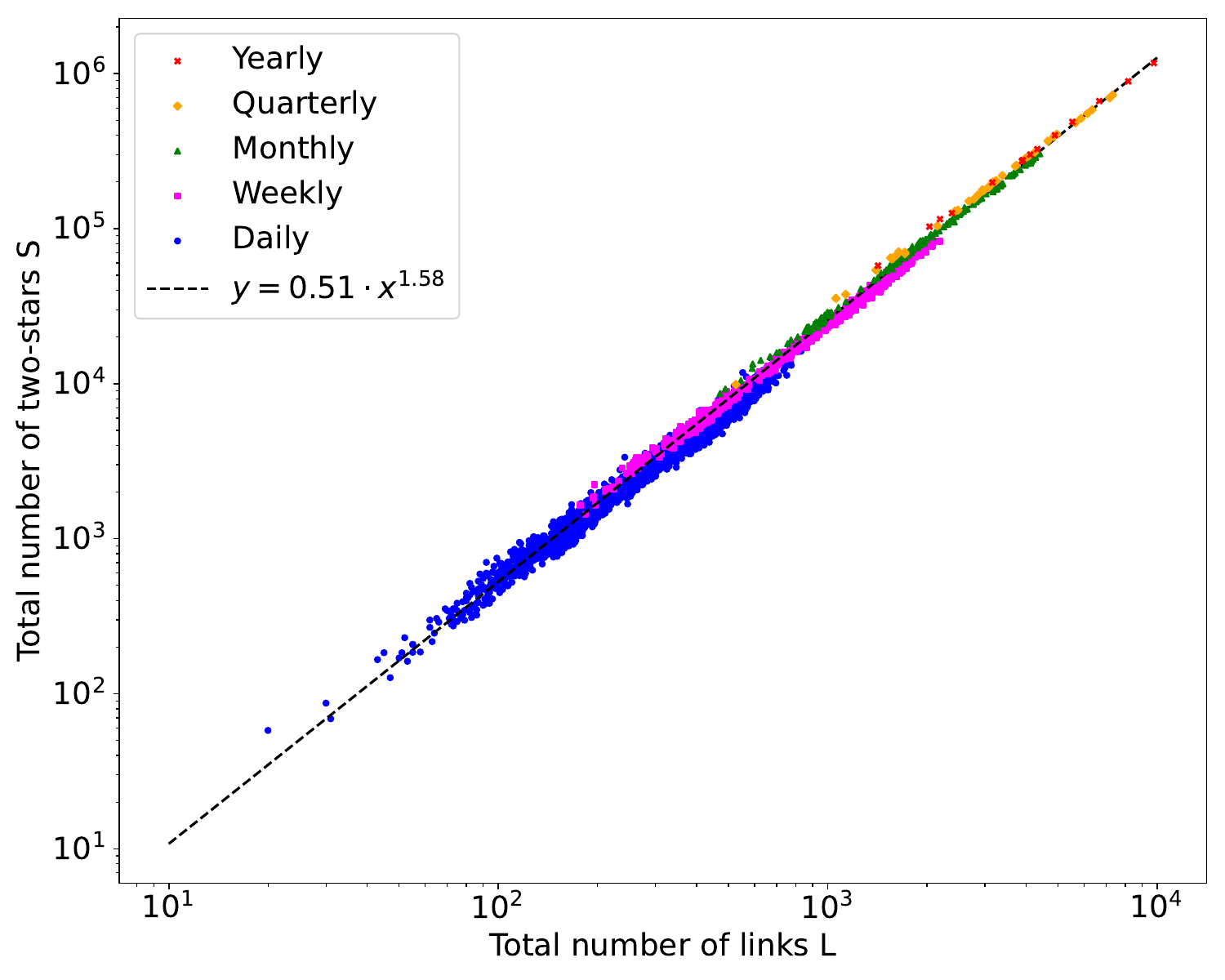}
\par\textbf{(b)}
\end{minipage}
\caption{\textbf{Validation and scaling of the fit2SM.} Panel~(a): solid lines represent the $z$-scores reading $z=(\mu-\lambda_1^\text{fit2SM})/\sigma$, while dashed lines represent the $z$-scores reading $z=(\mu-\lambda_1)/\sigma$. In both cases, $\mu$ and $\sigma$ have been calculated via the dcGM. The similarity of the two series of values indicate that the fit2SM represents a valid generative model (e.g., for early-warning signals detection) in case the true network topology is unavailable. The pre-crisis period $2005$-$2008$ and the crisis period $2008$-$2009$ are highlighted by shaded areas. Different colours indicate different time-scale: blue-daily, magenta-weekly, green-monthly, orange-quarterly, red-yearly. Trends have been smoothed via a rolling average over the points $[t-10,t+10]$ for the daily aggregation, over the points $[t-7,t+7]$ for the weekly aggregation, over the points $[t-5,t+5]$ for the monthly aggregation, over the points $[t-3,t+3]$ for the quarterly aggregation. Panel~(b): irrespectively from the aggregation level at which eMID is considered, a relationship like $S=aL^b$ seems to hold true, the average parameters reading $a\simeq0.51$ and $b\simeq1.58$. Each point is a network at a given level of aggregation.}
\label{fig:6}
\end{figure*}

A more explicit test concerns the accuracy of the fit2SM in reproducing the empirical spectral radius of the binary, undirected version of eMID across the years $1999$-$2012$: as the system we are considering cannot be expected to perfectly align with the Chung-Lu model, deviations from the previous results are expected; as the calculation of the mean and standard deviation of the absolute error $|\lambda_1^\text{model}-\lambda_1^\text{obs}|$ over the $M=10^3$ configurations drawn from the snapshot-specific ensemble of each model reveals, in fact, one finds that

\begin{itemize}
\item at the daily time-scale, the fit2SM achieves the smallest average error (i.e., $0.79\pm 0.51$), performing better than the UBCM and the dcGM in $94.00\%$ and $97.60\%$ of the snapshots, respectively;
\item at the weekly time-scale, the fit2SM achieves the smallest average error (i.e., $0.97\pm 0.66$), performing better than the UBCM and the dcGM in $82.40\%$ and $98.70\%$ of the snapshots, respectively;
\item at the monthly time-scale, the fit2SM achieves the smallest average error (i.e., $0.71\pm 0.66$), performing better than the UBCM and the dcGM in $64.80\%$ and $74.50\%$ of the snapshots, respectively;
\item at the quarterly time-scale, the UBCM achieves the smallest average error (i.e., $0.43\pm 0.21$), the fit2SM performing better than the dcGM in $67.30\%$ of the snapshots but performing better than the UBCM only in $32.70\%$ of the snapshots;
\item at the yearly time-scale, the UBCM achieves the smallest average error (i.e., $0.61\pm 0.57$), the fit2SM performing better than the dcGM in $100.00\%$ of the snapshots but performing better than the UBCM only in $14.30\%$ of the snapshots.
\end{itemize}

A related comparison is depicted in panel~(b) of fig.~\ref{fig:3}, illustrating the distribution - summarized by the violin plot - of the relative error $(\lambda_1^\text{model}-\lambda_1^\text{obs})/\lambda_1^\text{obs}$ in reproducing the empirical spectral radius, across snapshots and temporal aggregations. It is worth noticing that the occasional, superior performance of the UBCM on the densest snapshots comes at the cost of taking the entire degree sequence as input; the fit2SM, on the other hand, solely relies upon two, global parameters, a feature making such a model substantially less demanding, in terms of informational requirements and numerical complexity.

\subsection{Reproducing a network structure}

A compact indicator of a model's performance in reconstructing a network structure is provided by the Bayesian Information Criterion (BIC), reading

\begin{equation}
\text{BIC}=c\ln V-2\mathcal{L},
\end{equation}
where $V=N(N-1)/2$ accounts for the system size, the number of parameters amounts at $c=N$ for the UBCM, $c=1$ for the dcGM, and $c=2$ for the fit2SM, and the log-likelihood reads

\begin{equation}
\mathcal{L}=\ln P(\mathbf{A})=\sum_i\sum_{j(>i)}[a_{ij}\ln p_{ij}+(1-a_{ij})\ln(1-p_{ij})]
\end{equation}
for all of them, with

\begin{equation}
p_{ij}^\text{UBCM}=\frac{x_ix_j}{1+x_ix_j},
\end{equation}

\begin{equation}
p_{ij}^\text{dcGM}=\frac{zs_is_j}{1+zs_is_j},
\end{equation}
and

\begin{equation}
p_{ij}^\text{fit2SM}=\frac{zs_is_jy^{\kappa_i^{(0)}+\kappa_j^{(0)}}}{1+zs_is_jy^{\kappa_i^{(0)}+\kappa_j^{(0)}}}.
\end{equation}

Testing the three models above on eMID reveals systematic, aggregation-dependent differences. To present the full batch of results in a compact and directly comparable way, fig.~\ref{fig:4} illustrates the distributions of

\begin{equation}\label{eq:bic}
\delta\text{BIC}_{\text{model}}=\frac{\text{BIC}_{\text{model}}-\text{BIC}_{\text{min}}}{\text{BIC}_{\text{model}}},
\end{equation}
where $\text{BIC}_{\text{min}}=\min_{\text{model}}\{\text{BIC}_{\text{model}}\}$; while a null value of such an index identifies the best-performing model on the considered snapshot, the larger its value, the worse the performance of the corresponding model. As fig.~\ref{fig:4} shows, our canonical, non-linear model

\begin{itemize}
\item steadily outperforms both the dcGM and the (much) more demanding UBCM at the daily and weekly time-scales, i.e., on sparser configurations (in fact, $\text{BIC}_{\text{fit2SM}}<\text{BIC}_{\text{dcGM}}$ on $90.76\%$ of daily snapshots and $97.63\%$ of weekly snapshots; $\text{BIC}_{\text{fit2SM}}<\text{BIC}_{\text{UBCM}}$ on $99.97\%$ of daily snapshots and $100.00\%$ of weekly snapshots);
\item competes with the dcGM at the monthly time-scale (in fact, $\text{BIC}_{\text{fit2SM}}<\text{BIC}_{\text{dcGM}}$ on $64.24\%$ of monthly snapshots and $\text{BIC}_{\text{fit2SM}}<\text{BIC}_{\text{UBCM}}$ on $33.94\%$ of monthly snapshots but $\text{BIC}_{\text{fit2SM}}<\text{BIC}_{\text{dcGM}}$ on $100.00\%$ of monthly snapshots and $\text{BIC}_{\text{fit2SM}}<\text{BIC}_{\text{UBCM}}$ on $97.78\%$ of monthly snapshots since 2009);
\item compete with the dcGM at the quarterly time-scale during the last snapshots of our dataset (in fact, $\text{BIC}_{\text{fit2SM}}<\text{BIC}_{\text{dcGM}}$ on $55.56\%$ of quarterly snapshots but $\text{BIC}_{\text{fit2SM}}<\text{BIC}_{\text{UBCM}}$ on $11.11\%$ of quarterly snapshots since 2009).
\end{itemize}

Overall, these findings suggest the presence of a dependency between the performance of the fit2SM and the network link density, i.e., the sparser the network, the better its performance. To test such an hypothesis, let us scatter the index of relative performance

\begin{equation}
\Delta\text{BIC}_{\text{model}}=\frac{\text{BIC}_{\text{model}}-\text{BIC}_\text{fit2SM}}{\text{BIC}_{\text{model}}}
\end{equation}
versus the relative error in reproducing the total number of two-stars

\begin{equation}
\Delta S_{\text{model}}=\frac{\langle S\rangle_{\text{model}}-\langle S\rangle_\text{fit2SM}}{\langle S\rangle_{\text{model}}}=\frac{\langle S\rangle_{\text{model}}-S^*}{\langle S\rangle_{\text{model}}}
\end{equation}
for both the UBCM and the dcGM. As fig.~\ref{fig:5} reveals, whenever $\Delta S_{\text{model}}$ is large - typically the case for sparser configurations - model selection favours the fit2SM.

\subsection{The fit2SM as a generative model}

On the basis of the aforementioned results, let us employ the fit2SM as a generative model and test the accuracy of the dcGM in recovering the early warning signals (EWS) shown in~\cite{macchiati2025spectral}: to this aim, let us compute the $z$-score reading $z=(\mu-\lambda_1^\text{fit2SM})/\sigma$, where $\mu$ and $\sigma$ are the expected value and standard deviation of the dcGM estimate computed over the eMID surrogate generated by the fit2SM. As panel~(a) of fig.~\ref{fig:6} shows, the $z$-score induced by the fit2SM closely mirrors the one evaluated with respect to the true network topology for all time-scale, a result confirming the robustness of the fit2SM as a reference model for EWS detection. Stated otherwise, in case the true network topology is not available, the fit2SM proxies quite effectively the ground truth.

\section{Discussion}

After reviewing a number of negative results about the possibility of defining an ERG constraining both the degree sequence and the total number of two-stars, we propose a minimal, canonical model capable of reproducing the variance of a degree distribution while accounting for the nodes heterogeneity: in order to achieve such a goal, a fitness-induced variant of the 2SM, named fit2SM (i.e., a `softened' non-linear ERG whose definition is inspired to its fitness-based, linear counterpart), has been employed.

Besides achieving a large accuracy in reproducing a network structure at different levels, the fit2SM seems also to mitigate a problem affecting the dcGM, i.e. that of returning sampled configurations with too many isolated nodes: in fact, $\langle N_{\text{fit2SM}}^0\rangle<\langle N_{\text{dcGM}}^0\rangle$ on $97.63\%$ of daily snapshots, $98.61\%$ of weekly snapshots, and $67.88\%$ of monthly snapshots; interestingly, $\langle N_{\text{fit2SM}}^0\rangle<\langle N_{\text{UBCM}}^0\rangle$ on $58.79\%$ of monthly snapshots and $38.10\%$ of quarterly snapshots but $\langle N_{\text{fit2SM}}^0\rangle<\langle N_{\text{UBCM}}^0\rangle$ on $66.67\%$ of monthly snapshots and $55.56\%$ of quarterly snapshots since 2009 (see also Appendix~\hyperlink{AppF}{F}).

A last comment concerns the availability of the information about the total number of two-stars: this is rarely the case. As panel~(b) of fig.~\ref{fig:6} shows, however, it can be deduced from the one concerning the total number of links: irrespectively from the aggregation level, in fact, a relationship like $S=aL^b$ seems to hold true, the average parameters reading $a\simeq0.51$ and $b\simeq1.58$ (see also Appendix~\hyperlink{AppD}{D}). An analogous relationship holds true for the data concerning the yearly snapshots of the International Trade Network from 1990 to 2000~\cite{macchiati2025spectral}, the average parameters, now, reading $a\simeq0.44$ and $b\simeq1.61$.

Future research will explore non-linear models enforcing more complex patterns (e.g., the Strauss one, constraining the total number of triangles).

\section{Data availability}

The data supporting the findings of the present contribution are subject to proprietary restrictions and cannot be shared.

\section{Code availability}

The Python package named \texttt{fit2SM}, implementing the algorithms described in the main text, is available on PyPI and at the URL \url{https://github.com/mattiamarzi/fit2SM}.

\section{Acknowledgments}

MM, DG and TS acknowledge support from the project ‘SoBigData.it - Strengthening the Italian RI for Social Mining and Big Data Analytics’ - IR0000013 - CUP B53C22001760006, financed by European Union - Next Generation EU - National Recovery and Resilience Plan (Piano Nazionale di Ripresa e Resilienza, PNRR) - M4C2 I.3.1; DG and TS acknowledge support from the projects `Reconstruction, Resilience and Recovery of Socio-Economic Networks' RECON-NET EP\_FAIR\_005 - PE0000013 `FAIR', Funded by the European Union Next Generation EU, PNRR Mission 4 Component 2, Investment 3.1; `RE-Net: Reconstructing economic networks: from physics to machine learning and back' - 2022MTBB22, Funded by the European Union Next Generation EU, PNRR Mission 4 Component 2 Investment 1.1, CUP: D53D23002330006; `C2T - From Crises to Theory: towards a science of resilience and recovery for economic and financial systems' - P2022E93B8, Funded by the European Union Next Generation EU, PNRR Mission 4 Component 2 Investment 1.1, CUP: D53D23019330001. This work is also supported by the European Union - NextGenerationEU and funded by the Italian Ministry of University and Research (MUR) - National Recovery and Resilience Plan (Piano Nazionale di Ripresa e Resilienza, PNRR), project `A Multiscale integrated approach to the study of the nervous system in health and disease' (MNESYS, PE0000006, DN. 1553, 11/10/2022). We thank Giulio Cimini for fruitful and insightful discussions.

\section{Author contributions}

Study conception and design: MM, FG, DG, TS. Data collection: MM. Analysis and interpretation of results: MM, FG, DG, TS. Draft manuscript preparation: MM, FG, TS.

\section{Competing Interests}

The authors declare no competing interests.

\bibliography{references}

\clearpage

\onecolumngrid

\hypertarget{AppA}{}
\section*{APPENDIX A.\\DATA DESCRIPTION AND TEMPORAL AGGREGATIONS}\label{AppA}

We employ transaction-level records from the Electronic Market for Interbank Deposits (eMID), a screen-based market for unsecured deposits~\cite{macchiati2025spectral,IoriOvernightMoneyMarket2008,FingerFrickeLuxEMID2012}, restricting the analysis to the segment that represents the vast majority of the activity on such a market, i.e., the overnight one. Our sample spans trading days from January $1999$ to September $2012$: each trading day $d$ defines a weighted, directed matrix $\mathbf{V}(d)$ whose entry $v_{ij}(d)$ equals the total notional amount lent by bank $i$ to bank $j$ on day $d$; naturally, $v_{ii}(d)=0$.

Throughout the paper, the labels `daily', `weekly', `monthly', `quarterly' and `yearly' refer to calendar aggregations of daily records: more specifically, `weekly' refers to ISO calendar weeks (Monday to Sunday), `monthly' to calendar months, `quarterly' to standard quarters (Q1: Jan-Mar; Q2: Apr-Jun; Q3: Jul-Sep; Q4: Oct-Dec) and `yearly' to calendar years. Importantly, trading on eMID only occurs on business days, so that weekends and bank holidays do not contribute to observations: consequently, the number of trading days within a given calendar window is not constant across windows of the same kind; in practice, we aggregate over the set of trading days that are present in the raw records during a specific window. Let $\Delta_t$ denote the set of trading days belonging to the calendar window indexed by $t$ (week, month, quarter, or year). We, thus, aggregate weights according to

\begin{equation}
v_{ij}^{\Delta}(t)=\sum_{d\in\Delta_t}v_{ij}(d);
\end{equation}
as a last observation, let us stress that, for each window, we restrict the set of nodes to the banks that are active within that window, i.e., that are involved in at least one transaction during $\Delta_t$: the number of nodes, thus, becomes a time-dependent quantity.

Since the empirical records are weighted and directed, we construct an undirected exposure matrix by symmetrising the aggregated weights as

\begin{equation}
w_{ij}^{\Delta}(t)=v_{ij}^{\Delta}(t)+v_{ji}^{\Delta}(t),\quad i\neq j
\end{equation}
and define the corresponding binary adjacency matrix as

\begin{equation}
a_{ij}^{\Delta}(t)=\mathbbm{1}\left\{w_{ij}^{\Delta}(t)>0\right\},\quad i\neq j
\end{equation}
with $a_{ii}^{\Delta}(t)=0$. Naturally, we compute the node strengths from the underlying, symmetrized weights as

\begin{equation}
s_i^{\Delta}(t)=\sum_{j(\neq i)}w_{ij}^{\Delta}(t)
\end{equation}
and use them as the exogenous fitnesses informing both the dcGM and the fit2SM.

\clearpage

\hypertarget{AppB}{}
\section*{APPENDIX B.\\FROM LINEAR HAMILTONIANS TO SEPARABLE HAMILTONIANS}\label{AppB}

To ease the mathematical manipulations, let us focus on binary, undirected networks. Linear ERGs are described by Hamiltonians reading

\begin{equation}
H(\mathbf{A})=\sum_i\sum_{j(>i)}H_{ij}(a_{ij})=\sum_i\sum_{j(>i)}\theta_{ij}a_{ij},
\end{equation}
hence inducing models that can be factorized as

\begin{equation}
P(\mathbf{A})=\prod_i\prod_{j(>i)}p_{ij}^{a_{ij}}(1-p_{ij})^{1-a_{ij}}
\end{equation}
with

$$
p_{ij}=\frac{e^{-\theta_{ij}}}{1+e^{-\theta_{ij}}}=\frac{x_{ij}}{1+x_{ij}}.
$$

The related ensembles are easy to sample since $a_{ij}\sim\text{Ber}[p_{ij}]$ and $a_{ij}\perp\!\!\!\perp a_{kl}$, with $(i,j)\neq(k,l)$. Separable Hamiltonians, instead, read

\begin{align}\label{eq:17}
H(\mathbf{A})=\sum_i\sum_{j(>i)}H_{ij}(\mathbf{A})=\sum_i\sum_{j(>i)}a_{ij}\cdot f_{ij}(\mathbf{A},\theta_{ij}),
\end{align}
hence inducing models that can be factorized as well as

\begin{equation}
P(\mathbf{A})=\prod_i\prod_{j(>i)}p_{ij}^{a_{ij}}(1-p_{ij})^{1-a_{ij}}
\end{equation}
but with

\begin{equation}
p_{ij}=\frac{e^{-f_{ij}(\mathbf{P},\theta_{ij})}}{1+e^{-f_{ij}(\mathbf{P},\theta_{ij})}},
\end{equation}
where $\mathbf{P}=\langle\mathbf{A}\rangle$. The mean-field approximation consists precisely in replacing $\mathbf{A}$ with $\mathbf{P}$ in $f_{ij}$, in eq.~\ref{eq:17}.

\clearpage

\hypertarget{AppC}{}
\section*{APPENDIX C.\\THE TWO-STAR MODEL AND ITS DEGREE-CORRECTED VERSION}\label{AppC}

Let us provide a concrete example of the mean-field approximation above, by considering the so-called two-star model (2SM). It is defined by

\begin{align}
H_\text{2SM}(\mathbf{A})&=\theta L+\psi S\nonumber\\
&=\theta L+\psi\sum_i\sum_{l(>i)}V_{il}\nonumber\\
&=\theta L+\psi\sum_i\sum_{l(>i)}\sum_ja_{ij}a_{jl}\nonumber\\
&=\theta L+\frac{\psi}{2}\sum_i\sum_{l(\neq i)}\sum_ja_{ij}a_{jl}\nonumber\\
&=\theta L+\frac{\psi}{2}\sum_j\sum_i\sum_{l(\neq i)}a_{ij}a_{jl}\nonumber\\
&=\theta L+\frac{\psi}{2}\sum_j\sum_i\left[\sum_la_{ij}a_{jl}-a_{ji}\right]\nonumber\\
&=\theta L+\frac{\psi}{2}\left[\sum_i\sum_j\sum_la_{ij}a_{jl}-\sum_jk_j\right]\nonumber\\
&=\theta L+\frac{\psi}{2}\left[\sum_i\sum_ja_{ij}k_j-\sum_jk_j\right]\nonumber\\
&=\theta L+\frac{\psi}{2}\left[\sum_i\sum_{j(>i)}(a_{ij}k_j+a_{ji}k_i)-\sum_jk_j\right]\nonumber\\
&=\theta L+\frac{\psi}{2}\left[\sum_i\sum_{j(>i)}a_{ij}(k_i+k_j)-\sum_jk_j\right]\nonumber\\
&=\theta L+\frac{\psi}{2}\left[\sum_i\sum_{j(>i)}a_{ij}(k_i+k_j)\right]-\frac{\psi}{2}\sum_jk_j\nonumber\\
&=\theta L+\frac{\psi}{2}\left[\sum_i\sum_{j(>i)}a_{ij}(k_i+k_j)\right]-\psi L\nonumber\\
&=(\theta-\psi)L+\frac{\psi}{2}\left[\sum_i\sum_{j(>i)}a_{ij}(k_i+k_j)\right];
\end{align}
upon renaming $\theta-\psi$ as $\alpha$ and $\psi/2$ as $\beta$, one finds

\begin{align}
H_\text{2SM}(\mathbf{A})&=\alpha L+\beta\left[\sum_i\sum_{j(>i)}a_{ij}(k_i+k_j)\right]\nonumber\\
&=\alpha\sum_i\sum_{j(>i)}a_{ij}+\beta\left[\sum_i\sum_{j(>i)}a_{ij}(k_i+k_j)\right]\nonumber\\
&=\sum_i\sum_{j(>i)}a_{ij}[\alpha+\beta(k_i+k_j)]
\end{align}
and identifying $f_{ij}$ with $\alpha+\beta(k_i+k_j)$ leads to

\begin{equation}\label{eq:score_TwoStar}
p_{ij}^\text{2SM}=\frac{e^{-\alpha-\beta(k_i+k_j)}}{1+e^{-\alpha-\beta(k_i+k_j)}}=\frac{xy^{k_i+k_j}}{1+xy^{k_i+k_j}}.
\end{equation}

Since the information about the degrees is not supposed to be available, consistency requires that

\begin{equation}
p=\frac{xy^{2(N-1)p}}{1+xy^{2(N-1)p}};
\end{equation}
in case the information about the degrees is, instead, available, the Hamiltonian becomes 

\begin{equation}
H_\text{dc2SM}(\mathbf{A})=\sum_i\theta_ik_i+\psi S
\end{equation}
and induces the expression

\begin{equation}
p_{ij}^\text{dc2SM}=\frac{e^{-(\alpha_i+\alpha_j)-\beta(k_i+k_j)}}{1+e^{-(\alpha_i+\alpha_j)-\beta(k_i+k_j)}}=\frac{x_ix_jy^{k_i+k_j}}{1+x_ix_jy^{k_i+k_j}}.
\end{equation}

Let us now consider that

\begin{equation}
\sum_i\binom{\langle k_i\rangle}{2}=\frac{1}{2}\sum_i[\langle k_i\rangle^2-\langle k_i\rangle]=\frac{1}{2}\sum_i[\langle k_i^2\rangle-\text{Var}[k_i]-\langle k_i\rangle]=\langle S\rangle-\frac{1}{2}\sum_i\text{Var}[k_i].
\end{equation}
As a consequence, the expected number of two-stars reads

\begin{equation}
\langle S\rangle=\sum_i\binom{\langle k_i\rangle}{2}+\frac{1}{2}\sum_i\text{Var}[k_i].
\end{equation}

Since $S=\sum_i\binom{k_i}{2}$, requiring $\langle k_i\rangle=k_i$ leads one to obtain

\begin{equation}
\langle S\rangle=\sum_i\binom{k_i}{2}+\frac{1}{2}\sum_i\text{Var}[k_i]=S+\frac{1}{2}\sum_i\text{Var}[k_i]\geq S,
\end{equation}
the equivalence holding true in case $\text{Var}[k_i]=0$, $\forall\:i$ (i.e., either in the microcanonical case or in the canonical, deterministic case). If, instead, the following relationship holds true:

\begin{equation}
\label{condition_for_two_stars}
\sum_i\binom{\langle k_i\rangle}{2} < \sum_i\binom{k_i}{2}
\end{equation}
one finds that

\begin{align}
\langle S\rangle&=\sum_i\binom{\langle k_i\rangle}{2}+\frac{1}{2}\sum_i\text{Var}[k_i]<\sum_i\binom{k_i}{2}+\frac{1}{2}\sum_i\text{Var}[k_i]=S+\frac{1}{2}\sum_i\text{Var}[k_i],
\end{align}
i.e.,

\begin{equation}
\langle S\rangle<S+\frac{1}{2}\sum_i\text{Var}[k_i],
\end{equation}
or, even more explicitly,

\begin{equation}
\langle S\rangle-S<\frac{1}{2}\sum_i\text{Var}[k_i].
\end{equation}

In order for eq.~\ref{condition_for_two_stars} to hold true, the condition $\langle k_i\rangle<k_i$ must be verified for \emph{at least} one node: in other words, reproducing $S$ requires \emph{at least} one degree to be underestimated.

\clearpage

\begin{table*}[t!]
\begin{tabular}{|l|c|c|c|c|c|c|c|c|}
\hline
& $N$ & $L$ & $\langle L \rangle$ & $\text{MRE}_L$ & $S$ & $\langle S \rangle$ & $\text{MRE}_S$ & $\text{Time (s)}$ \\
\hline
\hline
e-MID 1999 - Full year & 215 & 9770 & 9770 & $9.08 \cdot 10^{-11}$ & $1168583$ & $1168583$ & $4.95 \cdot 10^{-11}$ & $5.69$ \\
\hline
e-MID 1999 - 2nd quarter & 200 & 6536 & 6536 & $3.43 \cdot 10^{-10}$ & $598078$ & $598078$ & $1.42 \cdot 10^{-10}$ & $2.23$ \\
\hline
e-MID 1999 - 5th month & 194 & 4153 & 4153 & $7.55 \cdot 10^{-10}$ & $271858$ & $271858$ & $1.96 \cdot 10^{-10}$ & $0.65$ \\
\hline
e-MID 1999 - 15th week & 191 & 1920 & 1920 & $1.42 \cdot 10^{-09}$ & $67109$ & $67109$ & $9.85 \cdot 10^{-11}$ & $0.41$ \\
\hline
e-MID 1999 - 160th day & 174 & 629 & 629 & $7.78 \cdot 10^{-10}$ & $10537$ & $10537$ & $4.57 \cdot 10^{-11}$ & $0.16$ \\
\hline
\hline
e-MID 2004 - Full year & 175 & 4327 & 4327 & $2.41 \cdot 10^{-10}$ & $325442$ & $325442$ & $1.17 \cdot 10^{-10}$ & $4.63$ \\
\hline
e-MID 2004 - 2nd quarter & 163 & 2948 & 2948 & $5.22 \cdot 10^{-10}$ & $170352$ & $170352$ & $1.99 \cdot 10^{-10}$ & $0.79$ \\
\hline
e-MID 2004 - 5th month & 152 & 1995 & 1995 & $1.01 \cdot 10^{-09}$ & $85170$ & $85170$ & $1.85 \cdot 10^{-10}$ & $0.36$ \\
\hline
e-MID 2004 - 15th week & 135 & 925 & 925 & $1.70 \cdot 10^{-09}$ & $21199$ & $21199$ & $9.68 \cdot 10^{-11}$ & $0.15$ \\
\hline
e-MID 2004 - 160th day & 120 & 407 & 407 & $1.85 \cdot 10^{-09}$ & $5014$ & $5014$ & $4.82 \cdot 10^{-10}$ & $0.10$ \\
\hline
\hline
e-MID 2012 - Full year & 97 & 1421 & 1421 & $5.85 \cdot 10^{-10}$ & $57671$ & $57671$ & $1.56 \cdot 10^{-10}$ & $0.41$ \\
\hline
e-MID 2012 - 2nd quarter & 85 & 912 & 912 & $5.70 \cdot 10^{-10}$ & $27417$ & $27417$ & $5.62 \cdot 10^{-11}$ & $0.19$ \\
\hline
e-MID 2012 - 5th month & 82 & 545 & 545 & $6.58 \cdot 10^{-10}$ & $10474$ & $10474$ & $9.89 \cdot 10^{-11}$ & $0.12$ \\
\hline
e-MID 2012 - 15th week & 66 & 238 & 238 & $1.14 \cdot 10^{-09}$ & $2859$ & $2859$ & $1.28 \cdot 10^{-10}$ & $0.07$ \\
\hline
e-MID 2012 - 160th day & 54 & 103 & 103 & $2.36 \cdot 10^{-09}$ & $531$ & $531$ & $8.16 \cdot 10^{-10}$ & $0.02$ \\
\hline
\end{tabular}
\caption{\label{tab:iterative} Performance of the fixed-point algorithm to solve the systems of equations defining the fit2SM on several snapshots of eMID ($N$ is the total number of nodes, $L$ is the total number of links and $S$ is the total number of two-stars).}
\end{table*}

\hypertarget{AppD}{}
\section*{APPENDIX D.\\THE FITNESS-INDUCED TWO-STAR MODEL AND ITS NUMERICAL RESOLUTION}\label{AppD}

In order to solve the system of equations defining the fit2SM, an appropriate vector of initial conditions needs to be chosen. In order to solve the dcGM, we have chosen $z_0=1$; in order to solve the fit2SM, we have chosen $\kappa_i^{(0)}=\kappa_i^{\text{dcGM}}$, $z_0=z_\text{dcGM}$, and $y_0=1$. As a stopping criterium, we have adopted a condition on the infinite norm of the vector of differences between the values of the parameters at subsequent iterations, i.e., $\max\{|\Delta z|,\:|\Delta y|\}<10^{-12}$. The accuracy of our method in estimating the constraints has been evaluated by computing the \textit{maximum relative errors} defined as $\text{MRE}_L=|L^*-\langle L\rangle|/L^*$ and $\text{MRE}_S=|S^*-\langle S\rangle|/S^*$. Table~\ref{tab:iterative} shows the time employed by our algorithm to converge as well as its accuracy in reproducing the constraints defining it. Overall, our method is fast and accurate: the numerical errors never exceed $O(10^{-1})$ and the time employed to achieve such an accuracy is always less than a minute.\\

\begin{figure*}[t!]
\centering
\begin{minipage}{0.49\linewidth}
\centering
\includegraphics[width=\linewidth]{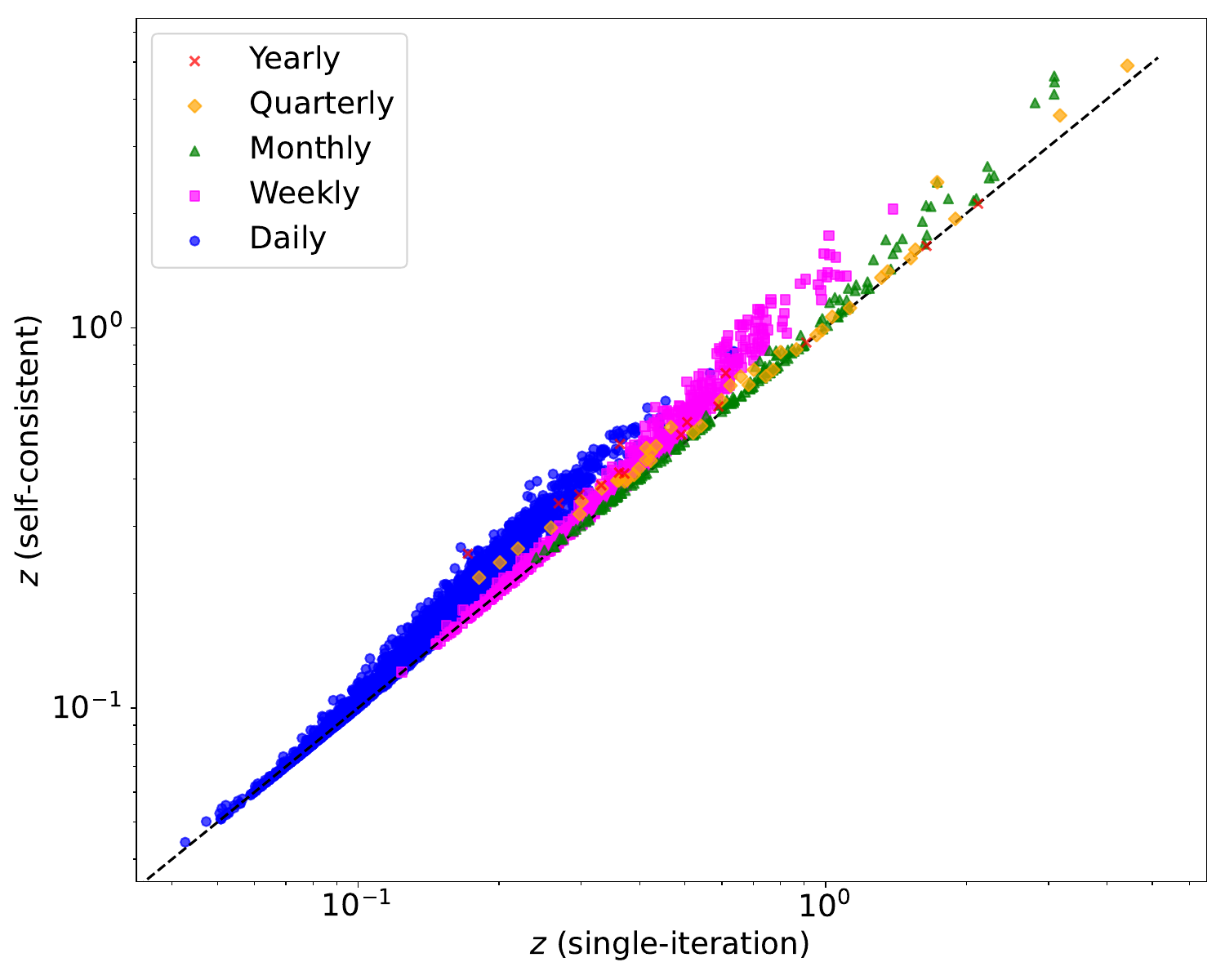}
\par\textbf{(a)}
\end{minipage}
\hfill
\begin{minipage}{0.49\linewidth}
\centering
\includegraphics[width=\linewidth]{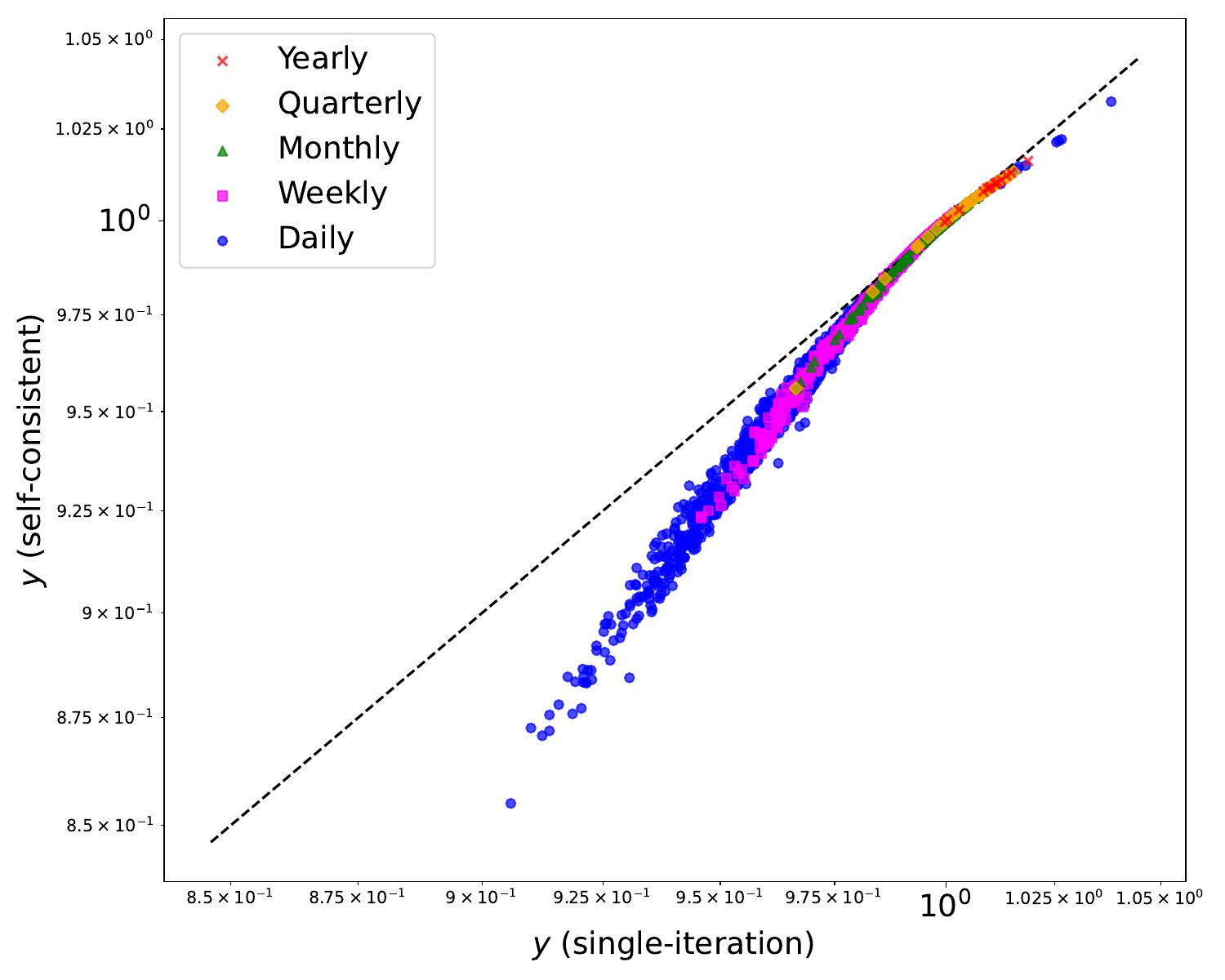}
\par\textbf{(b)}
\end{minipage}
\caption{\textbf{Single-iteration and self-consistent solutions of the fit2SM.} Panels~(a) and (b) compare the `single-iteration' solutions and the self-consistent ones for all the years of our dataset, at each time-scale.}
\label{fig:D1}
\end{figure*}

A natural question arises, i.e., does employing the `single-iteration' solution lead to significantly different results? To answer this question, we have compared the `single-iteration' (si) solutions with the self-consistent (sc) ones by calculating the maximum relative error concerning the average BIC values for each time-scale: one finds that $\left|\overline{\text{BIC}}_\text{sc}-\overline{\text{BIC}}_\text{si}\right|/\overline{\text{BIC}}_\text{sc}=0.0016$ at the daily time-scale, $\left|\overline{\text{BIC}}_\text{sc}-\overline{\text{BIC}}_\text{si}\right|/\overline{\text{BIC}}_\text{sc}=0.0011$ at the weekly time-scale, $\left|\overline{\text{BIC}}_\text{sc}-\overline{\text{BIC}}_\text{si}\right|/\overline{\text{BIC}}_\text{sc}=0.0003$ at the monthly time-scale, $\left|\overline{\text{BIC}}_\text{sc}-\overline{\text{BIC}}_\text{si}\right|/\overline{\text{BIC}}_\text{sc}=0.0008$ at the quarterly time-scale, and $\left|\overline{\text{BIC}}_\text{sc}-\overline{\text{BIC}}_\text{si}\right|/\overline{\text{BIC}}_\text{sc}=0.0008$ at the yearly time-scale (see also fig.~\ref{fig:D1}).\\

Let us now discuss an alternative way of determining the parameters of the fit2SM. First, let us write the log-likelihood as

\begin{equation}
\mathcal{L}_\text{fit2SM}=\sum_i\sum_{j(>i)}\left[a_{ij}\ln(zs_is_j y^{\kappa_i+\kappa_j})-\ln(1+zs_is_j y^{\kappa_i+\kappa_j})\right]
\end{equation}
and, then, maximize it with respect to $z$ and $y$. Upon doing so, we derive the two, non-linear, coupled equations reading

\begin{equation}
L^*=\sum_i\sum_{j(>i)}p_{ij}^\text{fit2SM}=\langle L\rangle
\end{equation}
and

\begin{equation}
\tau^*=\sum_i\sum_{j(>i)}a_{ij}(\kappa_i+\kappa_j)=\sum_i\sum_{j(>i)}p_{ij}^\text{fit2SM}(\kappa_i+\kappa_j)=\langle\tau\rangle;
\end{equation}
given that

\begin{equation}
S=\frac{1}{2}\sum_i\sum_{j(>i)}a_{ij}(k_i+k_j)-L=\frac{T}{2}-L,
\end{equation}
reproducing $L$ and (a proxy of) $T$ amounts at reproducing $L$ and (a proxy of) $S$. Since, however, we are interested in reproducing the empirical value of $S$, we have adopted the so-called \emph{method of moments}, imposing $\langle S\rangle=S^*$ in a (more) direct fashion.\\

In case the total number of two-stars were not directly accessible, one could exploit the relationship between $L$ and $S$, replacing $S^*$ with the value $S=aL^b$. For what concerns eMID, the fitted values read $a_{\text{daily}}\simeq0.36$, $b_{\text{daily}}\simeq1.59$; $a_{\text{weekly}}\simeq0.36$, $b_{\text{weekly}}\simeq1.61$; $a_{\text{monthly}}\simeq0.47$, $b_{\text{monthly}}\simeq1.59$; $a_{\text{quarterly}}\simeq0.67$, $b_{\text{quarterly}}=1.56$; and $a_{\text{yearly}}\simeq0.69$, $b_{\text{yearly}}=1.56$.

\clearpage

\begin{figure}[t!]
\centering
\begin{minipage}{0.49\textwidth}
\centering
\includegraphics[width=\linewidth]{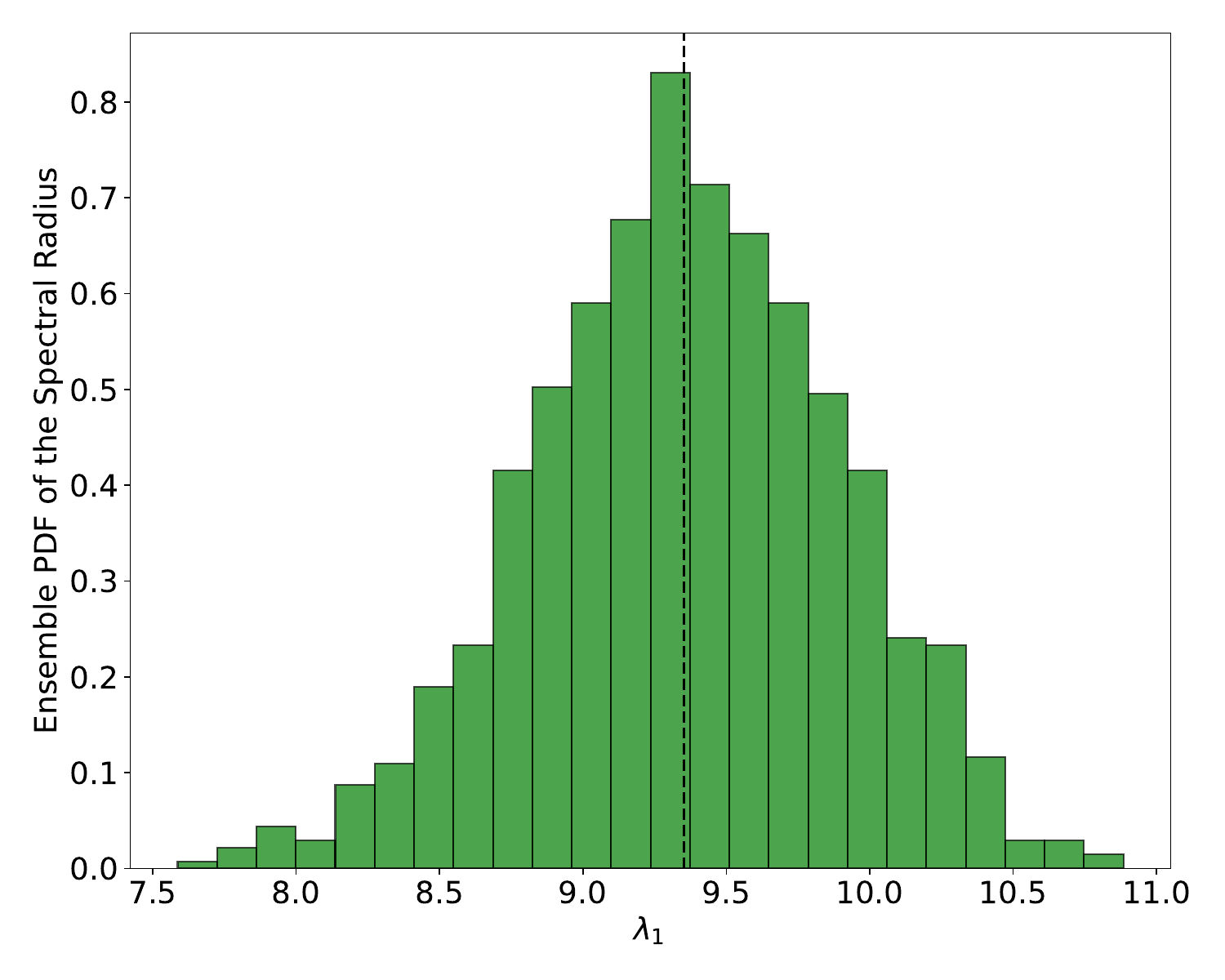}
\par\textbf{(a)}
\end{minipage}
\par\vspace{0.4em}
\begin{minipage}{0.32\textwidth}
\centering
\includegraphics[width=\linewidth]{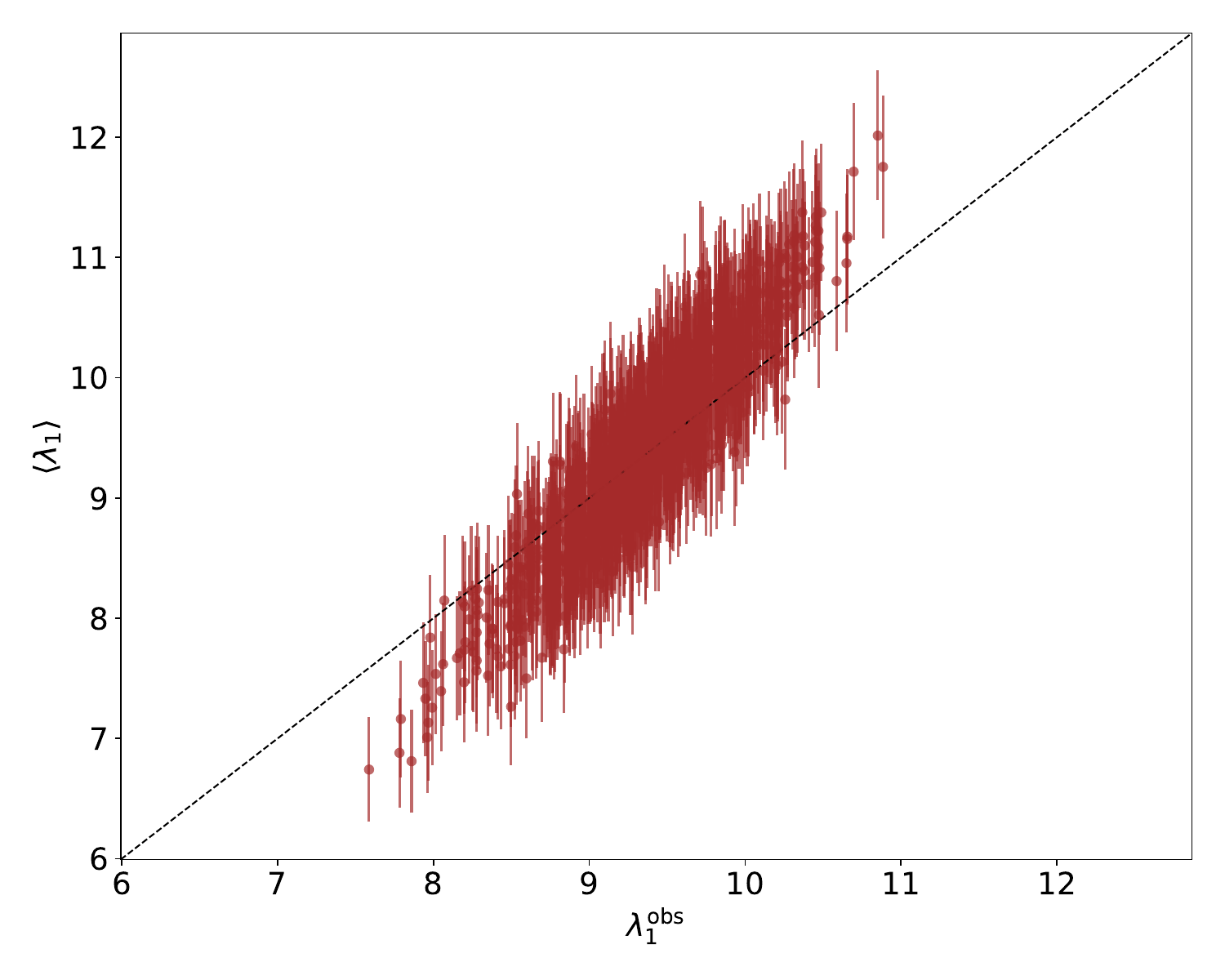}
\par\textbf{(b)}
\end{minipage}
\hfill
\begin{minipage}{0.32\textwidth}
\centering
\includegraphics[width=\linewidth]{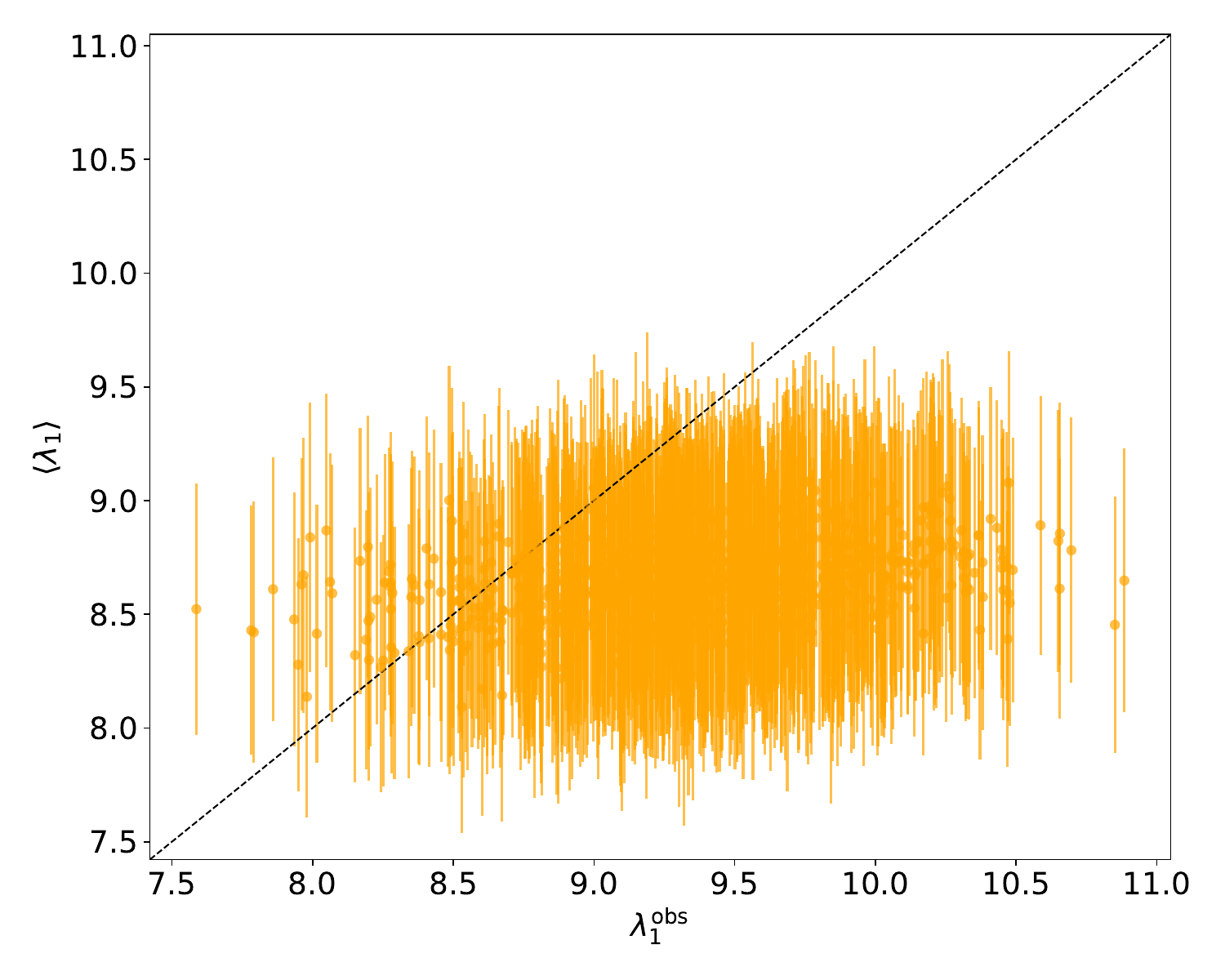}
\par\textbf{(c)}
\end{minipage}
\hfill
\begin{minipage}{0.32\textwidth}
\centering
\includegraphics[width=\linewidth]{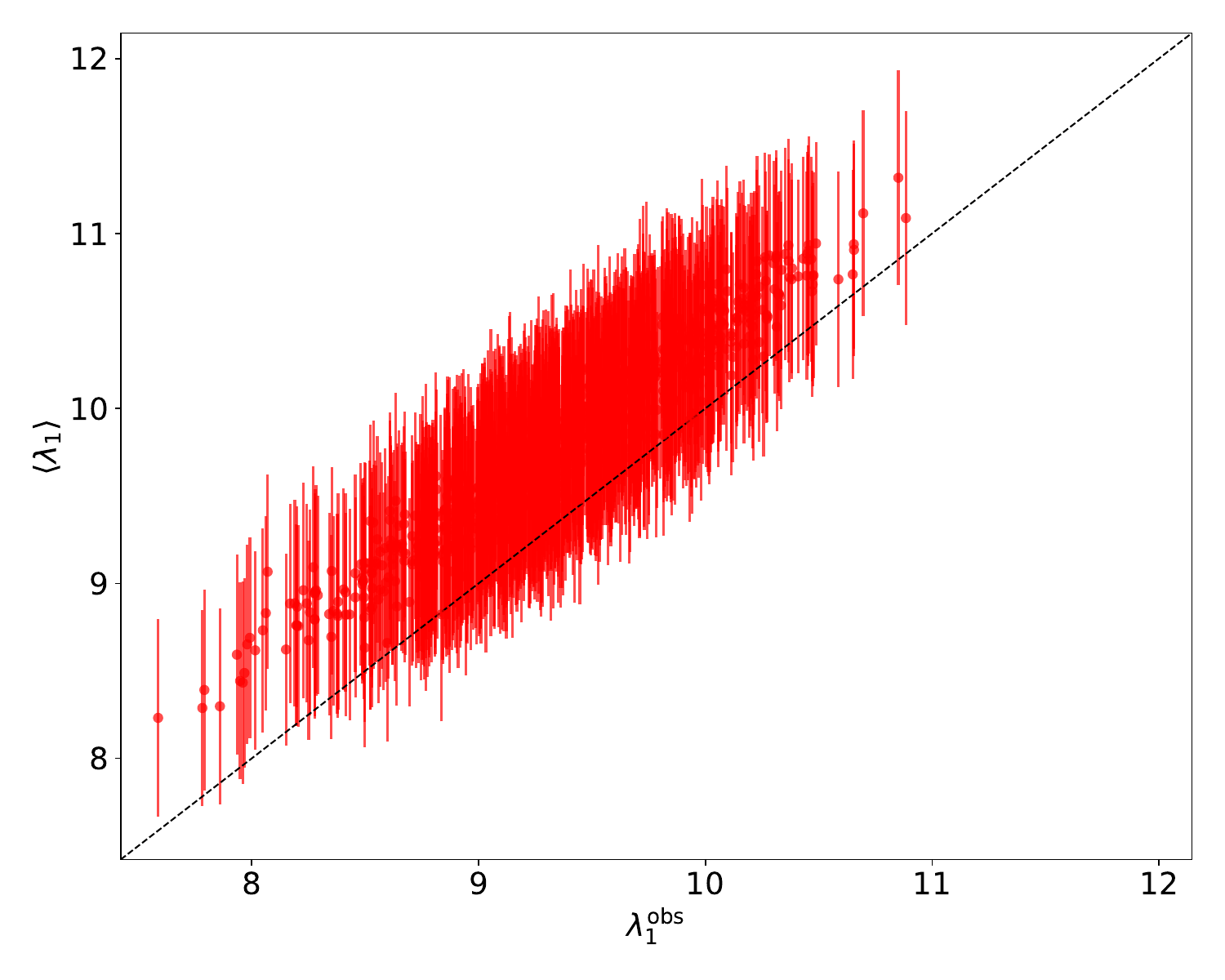}
\par\textbf{(d)}
\end{minipage}
\caption{\textbf{Estimating the spectral radius of synthetic configurations.} Panel~(a): distribution of the spectral radius induced by $M=10^3$ configurations sampled from the Chung-Lu model defined by the topology of the eMID snapshot corresponding to the day $2010$-$07$-$19$. Panels~(b--d): estimations of $\lambda_1$, obtained from each of the $M$, sampled configurations, scattered versus the corresponding empirical values, for the UBCM (red), the dcGM (yellow), and the fit2SM (brown). Vertical bars indicate the standard deviation accompanying the estimation carried out on the specific configuration. While the dcGM generally underestimates the spectral radius of the generative model and the UBCM overestimates it, the fit2SM captures it quite accurately.}
\label{fig:E1}
\end{figure}

\hypertarget{AppE}{}
\section*{APPENDIX E.\\SPECTRAL RADIUS}\label{AppE}

We now assess the ability of the considered models to reproduce the empirical value of a network spectral radius: to this aim, we pose ourselves within the controlled framework described in the main text.

Figure~\ref{fig:E1} shows the distribution of the spectral radius induced by the $M=10^3$ configurations sampled from the Chung-Lu model calibrated on the eMID snapshot corresponding to the day $2010$-$07$-$19$, the average $\langle\lambda_1\rangle\simeq 9.36$ being remarkably close to $\pi_1\simeq\overline{k^2}/\overline{k}=9.35$. Besides, we scatter the estimation of $\lambda_1$ obtained from each of the $M=10^3$, sampled configurations versus the corresponding empirical value: while the dcGM generally underestimates the latter, the UBCM overestimates it, as a consequence of its tendency to overestimate the variance of the degree distribution; the fit2SM, instead, displays the most accurate results.

\clearpage

\hypertarget{AppF}{}
\section*{APPENDIX F.\\PROBABILITY OF OBSERVING AT LEAST ONE ISOLATED NODE\\AND EXPECTED NUMBER OF ISOLATED NODES}\label{AppF}

Let us consider a sparse network. As such, dyads are independent and node $i$ is isolated with probability

\begin{equation}
q_i^0=P(k_i=0)=\prod_{j(\neq i)}(1-p_{ij})\simeq\prod_{j(\neq i)}e^{-p_{ij}}=e^{-\langle k_i\rangle};
\end{equation}
as a consequence, no node is isolated with probability

\begin{equation}
q^\bullet=\prod_i(1-q_i^0)=\prod_i\left[1-\prod_{j(\neq i)}(1-p_{ij})\right]\simeq\prod_i\left[1-\prod_{j(\neq i)}e^{-p_{ij}}\right]=\prod_i\left[1-e^{-\langle k_i\rangle}\right]
\end{equation}
and at least one node is isolated with probability

\begin{align}
q^0=1-q^\bullet=1-\prod_i(1-q_i^0)&=1-\prod_i\left[1-\prod_{j(\neq i)}(1-p_{ij})\right]\nonumber\\
&\simeq 1-\prod_i\left[1-\prod_{j(\neq i)}e^{-p_{ij}}\right]\nonumber\\
&=1-\prod_i\left[1-e^{-\langle k_i\rangle}\right].
\end{align}

\begin{figure*}[t!]
\centering
\begin{minipage}{0.49\linewidth}
\centering
\includegraphics[width=\linewidth]{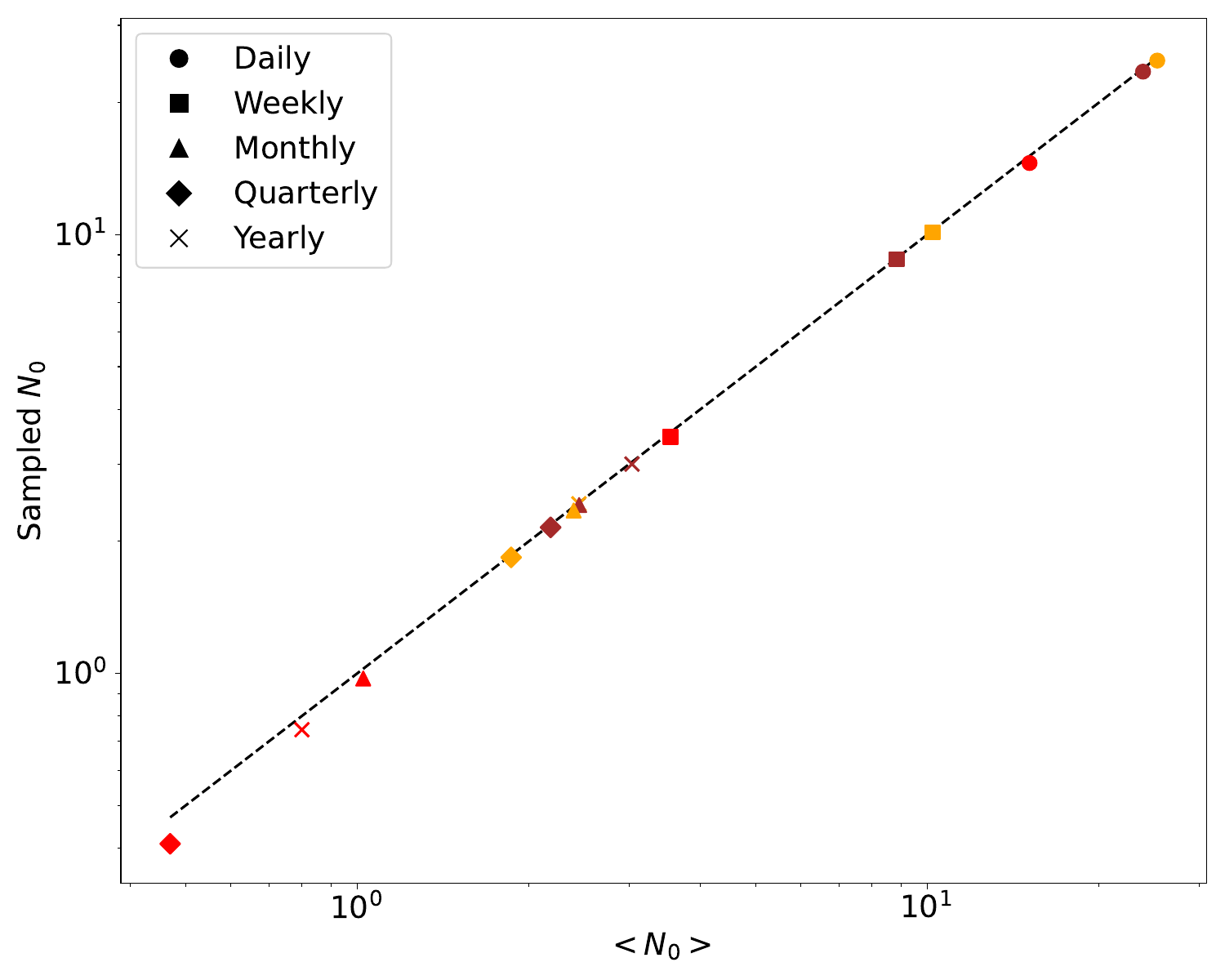}
\par\textbf{(a)}
\end{minipage}
\hfill
\begin{minipage}{0.49\linewidth}
\centering
\includegraphics[width=\linewidth]{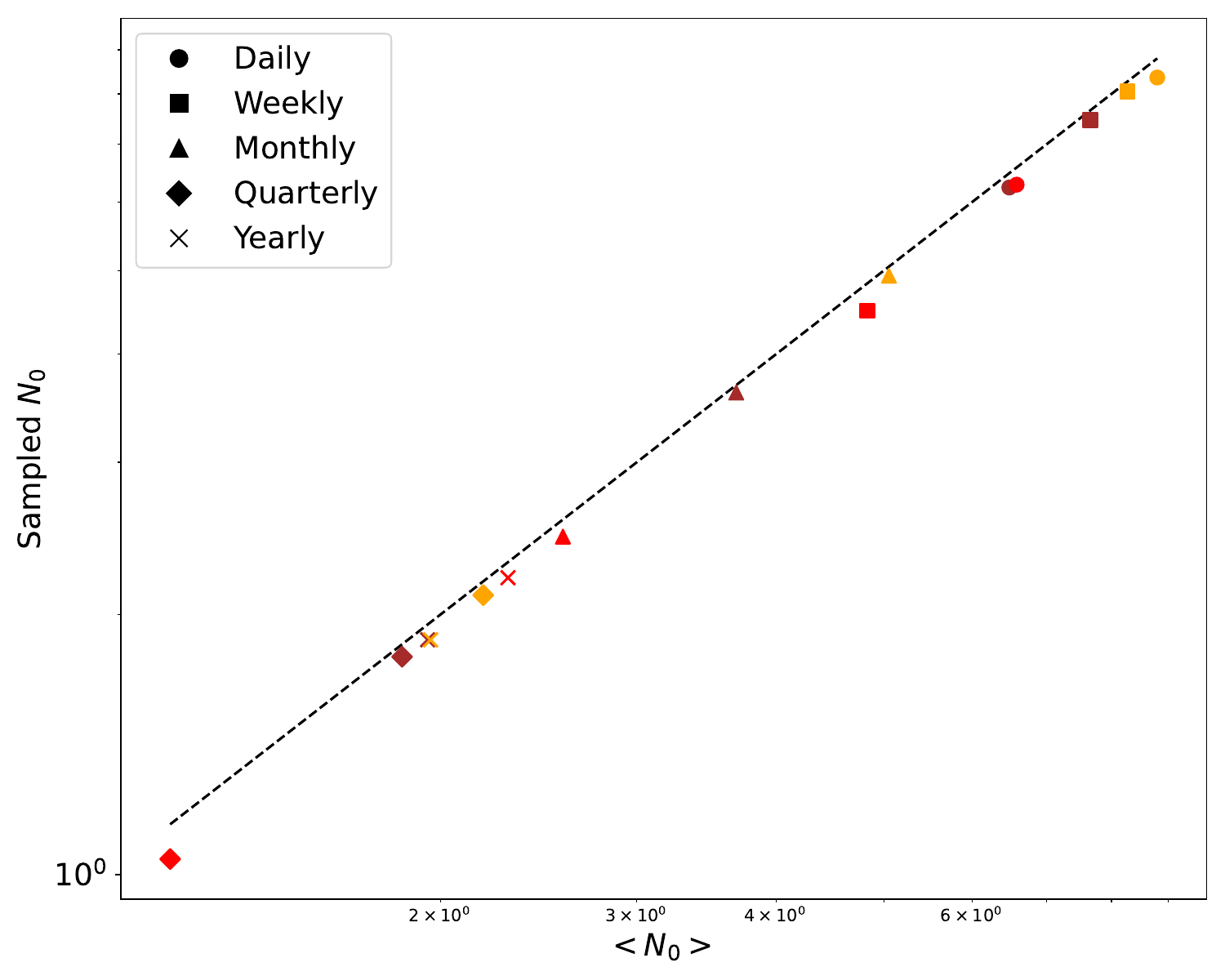}
\par\textbf{(b)}
\end{minipage}
\caption{\textbf{Estimating the number of isolated nodes.} Panels~(a) and (b) compare the ensemble averages of the total number of isolated nodes with their analytical counterparts for the weekly, monthly, quarterly, and yearly aggregation levels in 1999 and 2012, respectively: overall, the estimations provided by the 
UBCM (red), the dcGM (yellow), and the fit2SM (brown) are very accurate for each time-scale. The quarter, month, week, and day shown for each year correspond to those reported in Table~\ref{tab:iterative}.}
\label{fig:F1}
\end{figure*}

\begin{figure*}[t!]
\centering
\begin{minipage}{0.32\linewidth}
\centering
\includegraphics[width=\linewidth]{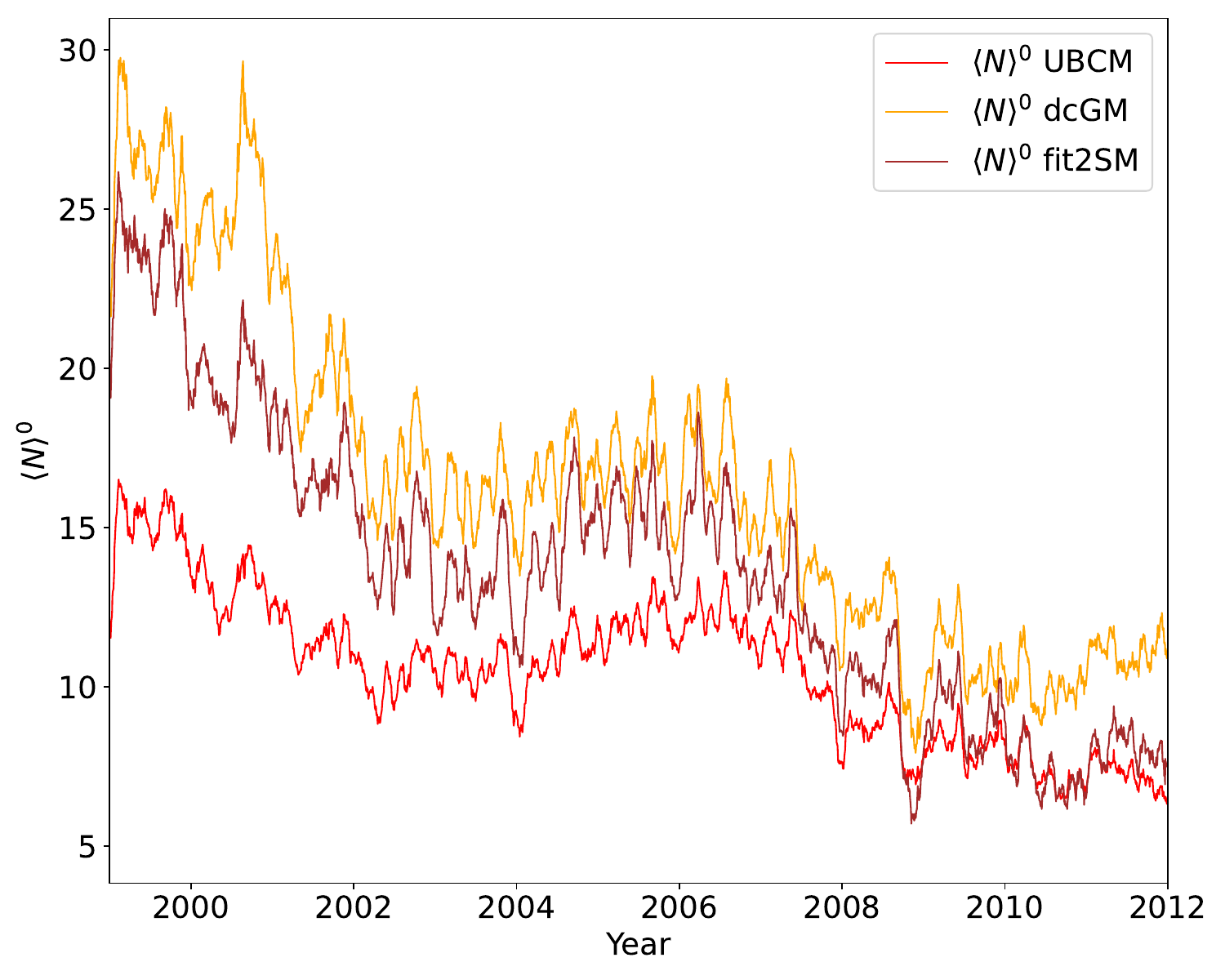}
\par\textbf{(a)}
\end{minipage}
\hfill
\begin{minipage}{0.32\linewidth}
\centering
\includegraphics[width=\linewidth]{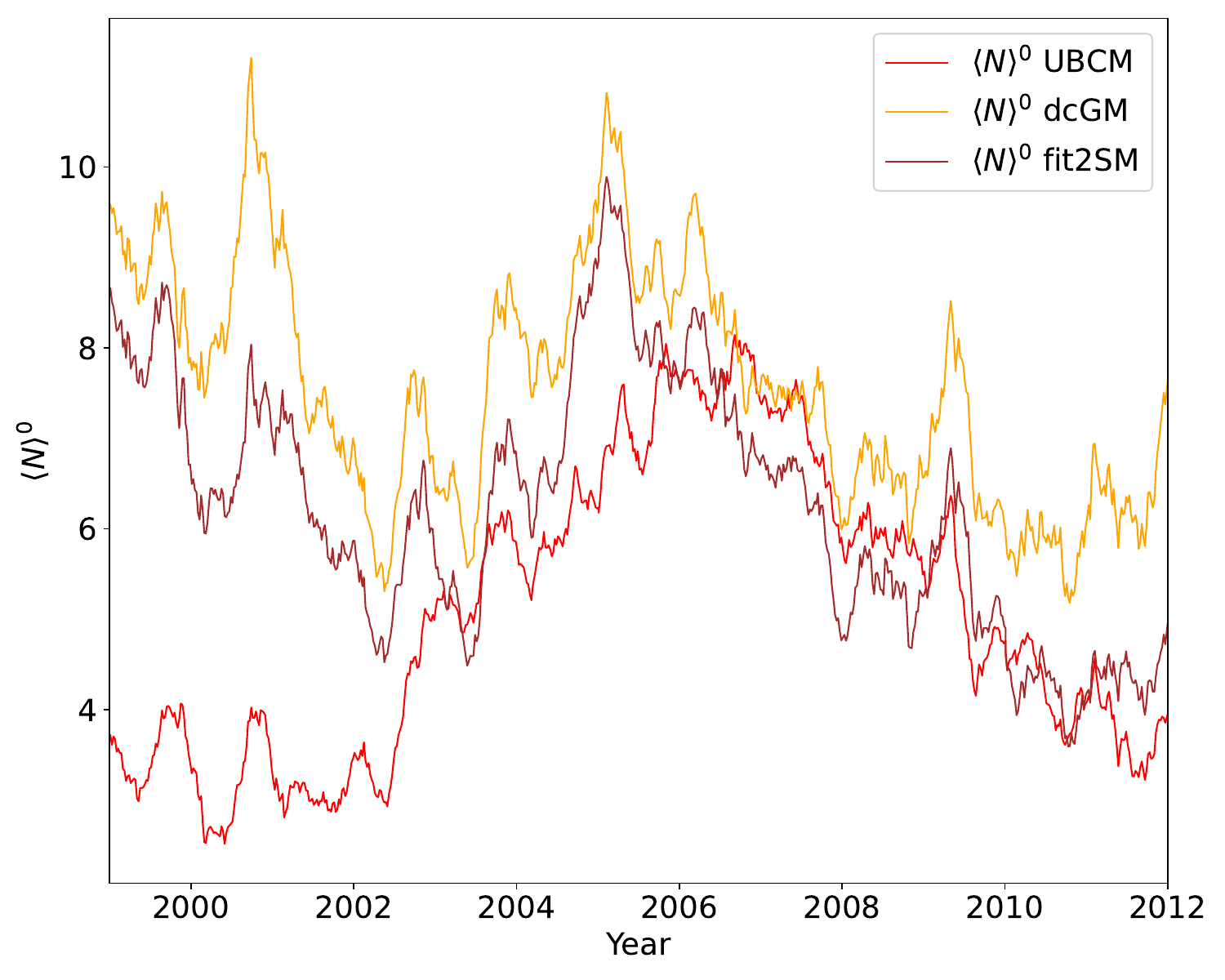}
\par\textbf{(b)}
\end{minipage}
\hfill
\begin{minipage}{0.32\linewidth}
\centering
\includegraphics[width=\linewidth]{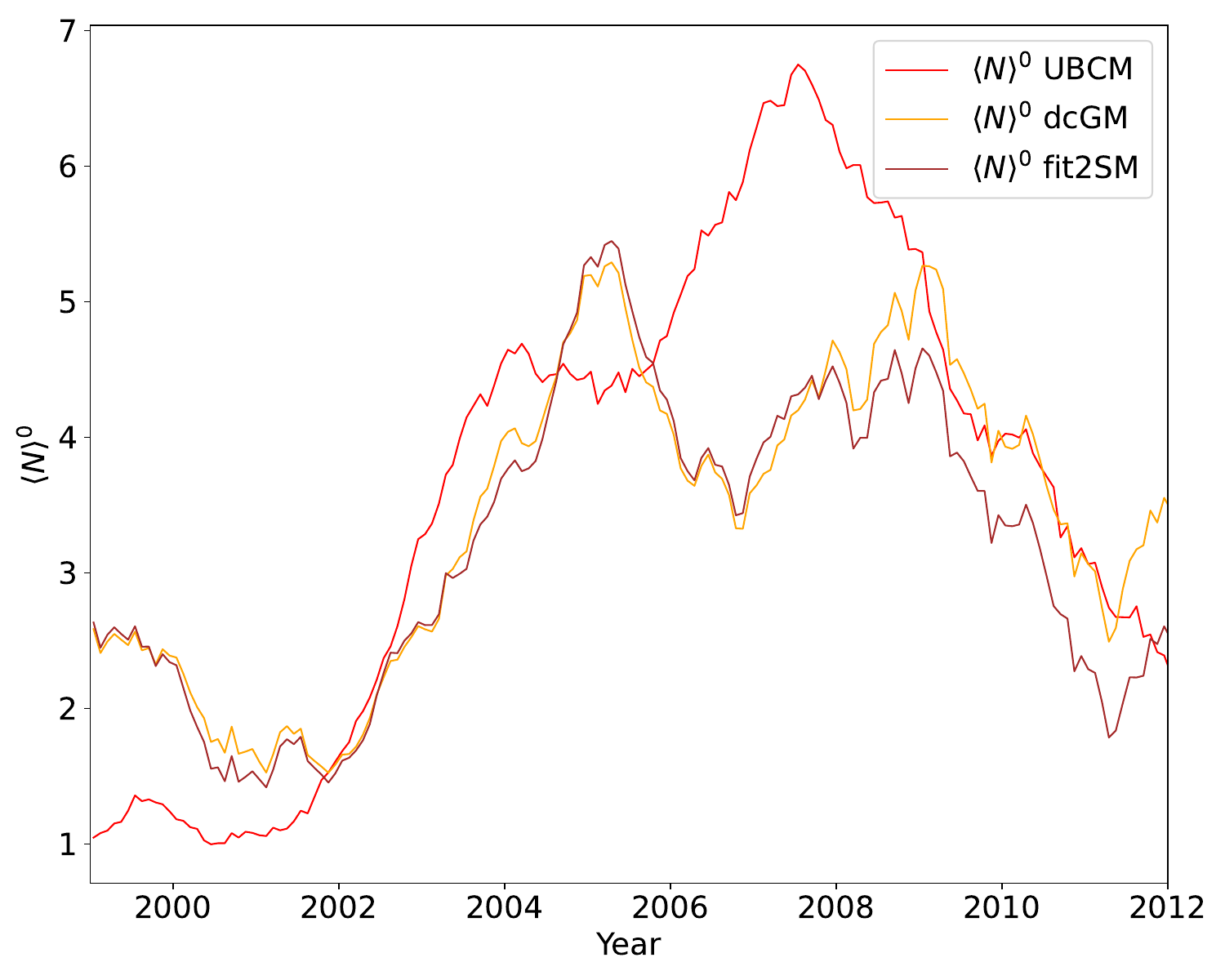}
\par\textbf{(c)}
\end{minipage}
\par\vspace{0.4em}
\begin{minipage}{0.32\linewidth}
\centering
\includegraphics[width=\linewidth]{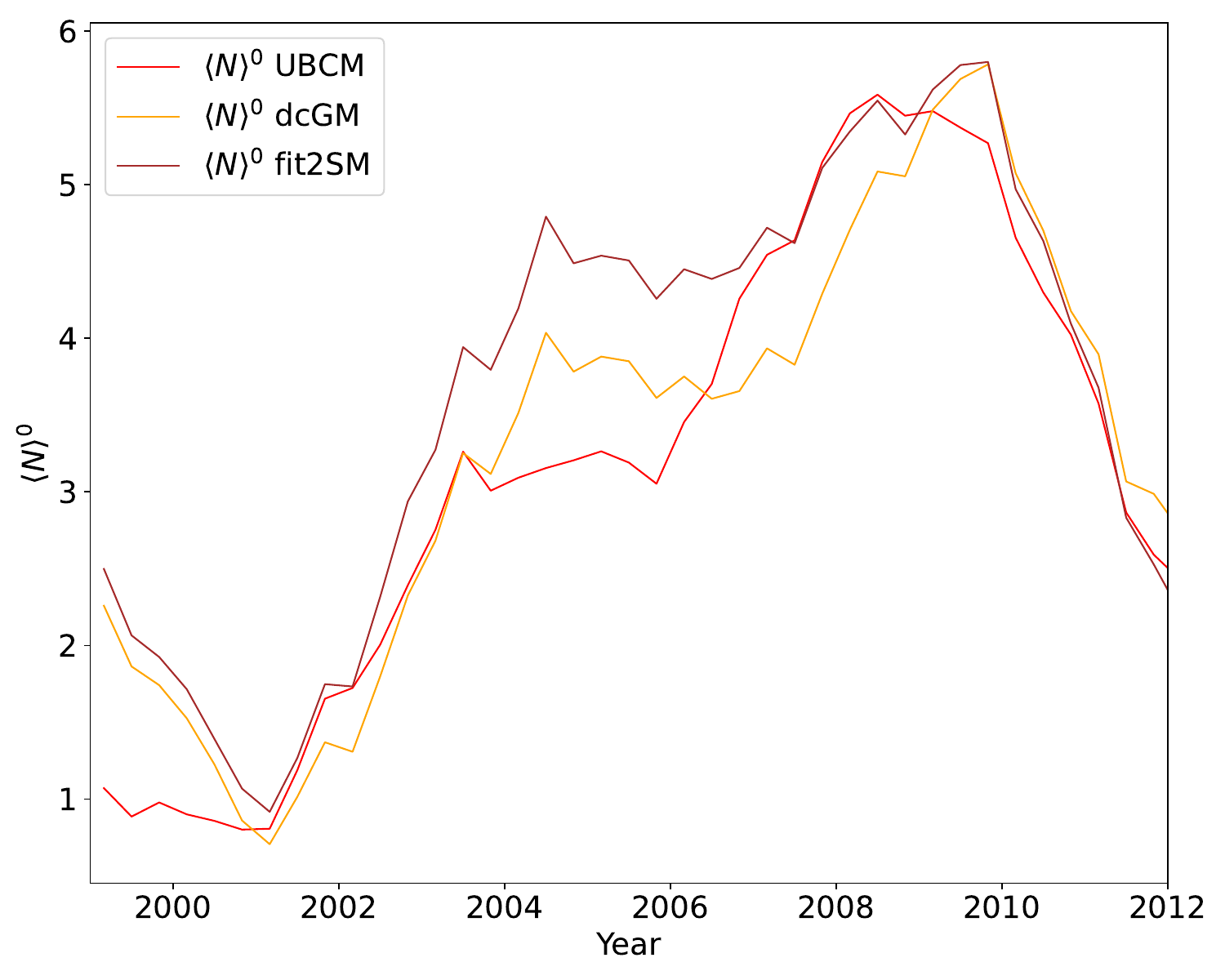}
\par\textbf{(d)}
\end{minipage}
\hspace{0.02\linewidth}
\begin{minipage}{0.32\linewidth}
\centering
\includegraphics[width=\linewidth]{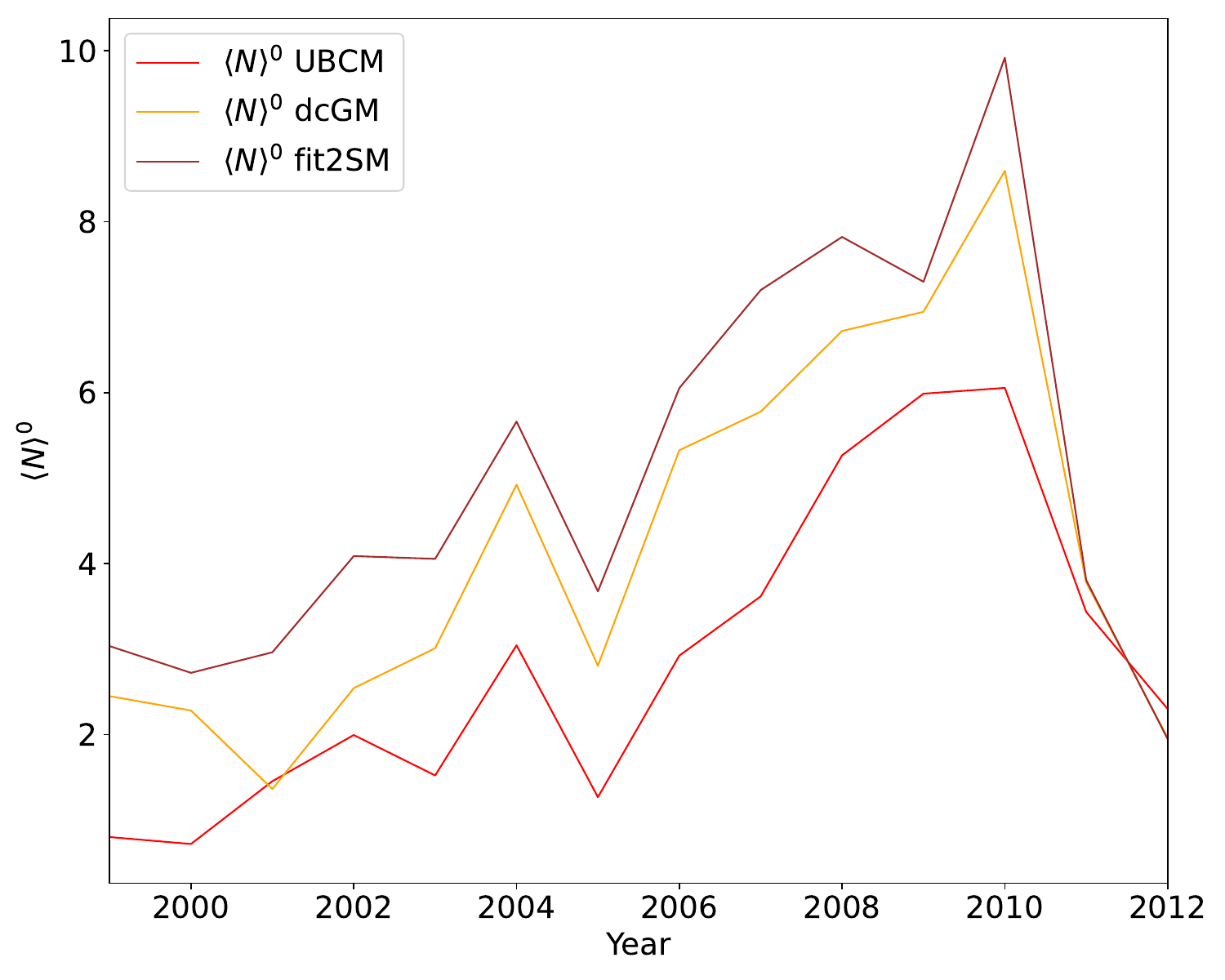}
\par\textbf{(e)}
\end{minipage}
\caption{\textbf{Estimating the number of isolated nodes across temporal aggregations.} Analysis of eMID at the daily, weekly, monthly, quarterly, and yearly time-scales (panels~(a--e), respectively). According to the expected number of isolated nodes, the fit2SM outperforms the dcGM at the daily and weekly time-scales and compete with the dcGM and the UBCM at the monthly and quarterly time-scales. Trends have been smoothed via a rolling average over the points $[t-10,t+10]$ for the daily aggregation, $[t-7,t+7]$ for the weekly aggregation, $[t-5,t+5]$ for the monthly aggregation, $[t-3,t+3]$ for the quarterly aggregation.}
\label{fig:F2}
\end{figure*}

Let us now compare 

\begin{equation}
q^0_\text{UBCM}=1-\prod_i\left[1-e^{-k_i}\right]\:\:\:\text{and}\:\:\:\langle N\rangle^0_\text{UBCM}=\sum_iq_i^0=\sum_ie^{-\langle k_i\rangle_\text{UBCM}}
\end{equation}
with

\begin{equation}
q^0_\text{dcGM}=1-\prod_i\left[1-e^{-\langle k_i\rangle_\text{dcGM}}\right]\:\:\:\text{and}\:\:\:\langle N\rangle^0_\text{dcGM}=\sum_iq_i^0=\sum_ie^{-\langle k_i\rangle_\text{dcGM}}
\end{equation}
and with

\begin{equation}
q^0_\text{fit2SM}=1-\prod_i\left[1-e^{-\langle k_i\rangle_\text{fit2SM}}\right]\:\:\:\text{and}\:\:\:\langle N\rangle^0_\text{fit2SM}=\sum_iq_i^0=\sum_ie^{-\langle k_i\rangle_\text{fit2SM}}
\end{equation}
across our temporal snapshots. Figure~\ref{fig:F1} compares the ensemble averages of the total number of isolated nodes with their analytical counterparts for the weekly, monthly, quarterly, and yearly aggregation levels in 1999 and 2012: overall, the estimations provided above are very accurate at each time-scale. Figure~\ref{fig:F2}, instead, compares the performances of the UBCM, the dcGM and the fit2SM in `producing' configurations being characterized by a certain number of isolated nodes: as we have already observed, the sparser the configuration, the better the performance of the fit2SM - in these cases, the fit2SM outperforms the dcGM and compete with the UBCM on at least a portion of the dataset, by allowing a smaller number of isolated nodes to appear on sampled configurations.

\end{document}